%% just a bit extra polish march 2026
%% for FocuStat pages and for arXiv hopefully 

%% topoffile 

%	This is `moderate_spurt.tex', and is intended to spurt even
%	more than `moderate_finish.tex'. 
%	Last changes: about January and February 1991.
%	Take heed: old x to new y, old beta to new gamma. Check out!

% Original version quickly written in April and May 1990. 
% Then some changes in January and February 1991.
% Then a ridiculous amount of time went by until I finally 
% 	computed my risk functions, and finished a good 
%	StatResReport version in August 1991.
% Then it spent time in the offices of StatScience and IntStatReview ...	
%	they liked it but wanted it in tradiotional research journals ... 
% Then a riduculous amount of time went by in my drawers ... 
% With renewed attempts at getting it published: 
%	Finally moving again, May 1993! Technometrics? 
%	Captain, Oh Captain: why does it take me three years from 
%	super start to real finish? 

\magnification\magstep1
\baselineskip14pt
\vsize24.0truecm 

\input miniltx
\input graphicx

\overfullrule=0pt

%% \input nils.sty

% This is "nils.sty" and proudly contains some standard macros 
% for my papers & reports &cetera. 
% Last touched on: December 1994: 
%       note that fermatboxsize should be 4.00pc usually in tech reporst,
%       but possible say 5.00pc in preliminary versions. 

% My intention is to make updates on the nils.sty file placed 
% in subdirectory `papers94', `papers95', and so on; 
% and to copy it into all the other required subdirectories. 
% For example: I have redesigned \hskipdistanceleft and \hskipdistanceright 
% ever so slightly ... 

% It includes standard definitions like \hatt, \eps etcetera,
% and \today, \footline, \square, \fermat, \ref, \beginverbatim ... |||

% These should be paper-dependent: 
% \magnification\magstep1 
% \vsize=24.0truecm
% \baselineskip13pt
% And then, typically, comes \input nils_macros.tex

% MATHEMATICAL MINI-MACROS: 
\def\hatt{\widehat}
\def\dell{\partial}
\def\tilda{\widetilde}
\def\eps{\varepsilon}

\def\half{\hbox{$1\over2$}}

\def\arr{\rightarrow}
\def\normal{{\cal N}}

\def\RR{\mathord{I\kern-.3em R}}
\def\PP{\mathord{I\kern-.3em P}}
\def\NN{\mathord{I\kern-.3em N}}
\def\ZZ{\mathord{I\kern-.3em Z}} 

\def\E{{\rm E}}
\def\d{{\rm d}}
\def\Pr{{\rm Pr}}
\def\mtrix{\pmatrix} 
\def\midd{{\,|\,}}
\def\subsection{\medskip}

% FONTS: 
\font\bigbf=cmbx12

\font\csc=cmcsc10

 at 10truept 
\font\smallrm=cmr8

% TODAY-MACRO:
\def\today{\number\day \space \ifcase\month\or
January\or February\or March\or April\or May\or June\or 
July\or August\or September\or October\or November\or December\fi  
\space \number\year}
% Or: can use `\def\today{\smallrm yesterday}', if I wish 

% REFERENCES-options:
% usually used in connexion with 
% \parindent0pt
% \baselineskip11pt
% \parskip3pt & \medskip to start the list off. 

% LaTeX references, according to someone
   
% TeX references, according to someone
\def\ref#1{{\noindent\hangafter=1\hangindent=20pt
  #1\smallskip}}          
% my usual preference 

% ATHLETE'S FOOTLINE: 
% \def\quotationone{\smallcyr Slava Trudu}
% \def\quotationtwo{\smallcyr Krasota Spaset Mir} 
\def\quotationone{\smallrm Where there is a Will}
\def\quotationtwo{\smallrm There is a Won't}
% or something else 
% \footline={\quotation\hfil{\rm\the\pageno}\hfil
%               {\smallrm\today}}
\def\hskipdistanceleft{\hskip-3.5pt}
\def\hskipdistanceright{\hskip-2.0pt}
\footline={{
\ifodd\count0
        {\hskipdistanceleft\quotationone\phantom{\smallrm\today}
                \hfil{\rm\the\pageno}\hfil
         \phantom{\quotationone}{\smallrm\today}\hskipdistanceright}
        \else 
        {\hskipdistanceleft\quotationtwo\phantom{\today}
                \hfil{\rm\the\pageno}\hfil
         \phantom{\quotationtwo}{\smallrm\today}\hskipdistanceright}
        \fi}}

% An easier option, with only one quotation defined
% (in which case the hskipdistances do not seem to be required?): 
% \footline={{
%        {\quotation\phantom{\smallrm\today}
%                \hfil{\rm\the\pageno}\hfil
%         \phantom{\quotation}{\smallrm\today}} }}
         
% SQUARE, BOXITs:
% should also call in my \boxit method...
\def\cstok#1{\leavevmode\thinspace\hbox{\vrule\vtop{\vbox{\hrule\kern1pt
        \hbox{\vphantom{\tt/}\thinspace{\tt#1}\thinspace}}
        \kern1pt\hrule}\vrule}\thinspace} % control sequence token
\def\square{\cstok{\phantom{$\cdot$}}} 
                % subsidiaert X; dette avgjoer stoerrelsen

% also check my concert programme notes ... 
% could also throw in \hskip-0.20truecm and the like ... 

\def\halmos{%
  \hbox{%
    \vrule width 0.5em height 0.5em depth 0pt
  }%
}

\def\square{\halmos}

% FERMAT: a margin-note trick 
\def\fermat#1{\setbox0=\vtop{\hsize4.00pc
        \smallrm\raggedright\noindent\baselineskip9pt
        \rightskip=0.5pc plus 1.5pc #1}\leavevmode
        \vadjust{\dimen0=\dp0
        \kern-\ht0\hbox{\kern-4.00pc\box0}\kern-\dimen0}}
% works
% \fermat{like this, you mean?}like this.
% should ideally use some `\eightpoint' universal command ... 
% at this stage only smallrm ok 
% This is nonrobust ... inconsequential behaviour ... 

\def\hsizeplusepsilon{14.25truecm} 
% in general, \def it properly as \hsize + 0.25truecm 
\def\fermatright#1{\setbox0=\vtop{\hsize4.00pc
        \smallrm\raggedright\noindent\baselineskip9pt
        \rightskip=0.5pc plus 1.5pc #1}\leavevmode
        \vadjust{\dimen0=\dp0
        \kern-\ht0\hbox{\kern\hsizeplusepsilon\box0}\kern-\dimen0}}

% \def\fermatquote{Cujus rei demonstrationem 
%        mira\-bilem sane detexi. Hanc marginis exiguitas non caperet.} 
% Full quote, 1637: 
% Cubum autem in duos cubos, aut quadratoquadratum in duos quadratoquadratos,
% et generaliter nullam in infinitum ultra quadratum potestatem
% in duos ejusdem nominis fas est dividere: 
% cujus rei demonstrationem mirabilem sane detexi. 
% Hanc marginis exiguitas non caperet

% VERBATIM: sort of works, not foolproof yet, 
% \beginverbatim & besides I & # $ am sometimes a fool |||
% In fact it gets confused when seeing an ordinary | . 
% \def\makeordinary{\catcode`\&=12 \catcode`\{=12
%       \catcode`\}=12 \catcode`\#=12 \catcode`\\=12 \catcode`\$=12 
%       \catcode`\_=12 \catcode`\^=12 \catcode`\%=12 \catcode`\~=12}
% \catcode`\|=\active
% \def\beginverbatim{\bgroup\makeordinary\obeylines\obeyspaces\tt%
%       \def|||{\egroup}}

% \ordinarypagenumbers
\def\today{May 1993}
\def\quotationone{\smallrm Nils Lid Hjort}
\def\quotationtwo{\smallrm Moderately misspecified models}

\def\delltheta{\hbox{$\dell\mu\over \dell\theta$}}

\def\dellgamma{\hbox{$\dell\mu\over \dell\gamma$}}

%{\bf When is it best to keep a moderately misspecified model?} 
\centerline{\bigbf Estimation in moderately misspecified models} 
% 	Nils: suboptimal title? 

\smallskip
\centerline{\bf Nils Lid Hjort } 

\smallskip
\centerline{\sl University of Oslo and University of Oxford}

\smallskip 
{{\smallskip\narrower\noindent\baselineskip12pt 
{\csc Abstract}. 
% new shortened version since I think of Technometrics ... 
Suppose data are fitted to some parametric model 
but that the true model happens to be one with an additional parameter.
When a parameter is to be estimated 
one can use likelihood estimation in the wider model or in the narrow model.  
Including the extra parameter in the model means less bias 
but larger sampling variability. 
Two basic questions are addressed in this article. 
(i) 
Just how much misspecification can the narrow model tolerate? 
In the context of a large-sample moderate misspecification framework 
we find a surprisingly simple, sharp, and general answer.
% surprisingly? 
% in the form of an explicit criterion for when narrow estimation 
% is more precise than wide estimation, valid for all estimands. 
There is effectively a `tolerance radius' around a given narrow model,
inside of which narrow estimation is more precise than wide estimation
for all estimands. 
This is computed in a selection of examples
% interesting? of course they are 
that also demonstrate the degree of robustness of 
important standard methods against moderate 
incorrectness of the model under which they are optimal. 
(ii) 
Are there other estimators that work well both 
under narrow and wide circumstances? 
We discuss several possibilities and propose some new procedures. 
All methods are compared in a broad large-sample performance study.
% This comparison can be carried out 
% rather generally and rather simply 
% due to a drastic reduction to a particular standard problem.
%	Lends support to classical methods
%	Nils: for he who increaseth his wisdom increaseth his grief.

\smallskip\noindent 
{\csc Key words:} \sl
Akaike criterion, 
% Bayes and empirical Bayes, 
compromise estimators, 
deliberate bias,
% ignorance is strength, 
misspecified mo\-del, 
% model choice criteria, 
parametric inference,
% performance study,
tolerance radius
\smallskip}}

\bigskip
{\bf 1. Introduction and motivating examples.} 
Our theme is moderately misspecified parametric models,
and we ask two main questions. 
The first is: Just how much misspecification 
can a given parametric model tolerate in a 
certain direction? 
More specifically, when is it advantageous to stick to the 
narrow model, even when it is incorrect? 
When will `narrow estimation' be more precise than
`wide estimation'?  
The second question is broader: 
Are there estimators that are about as good as 
the narrow estimator when the narrow model is correct,
and about as good as the wide estimator when the 
narrow model is incorrect? 
%	--- could point out that estimation is main concern --- 
We shall present a generous list of examples to
motivate the problems and precise formulations of them. 

\subsection
{\csc Example A.} 
Suppose data $Y_1,\ldots,Y_n$ come from a life distribution 
%with density $f$ 
on $[0,\infty)$ and that the median $\mu$
is to be estimated. If the density is the exponential 
$f(y)=\theta e^{-\theta y}$, then $\mu=\log 2/\theta$, and 
a natural estimator is $\hatt\mu_{\rm narr}=\log 2/\hatt\theta_{\rm narr}$,
where $\hatt\theta_{\rm narr}=1/\bar Y$ is the 
maximum likelihood (ML) estimator
in this narrow model. If it is suspected that the model could
deviate from simple exponentiality in direction of the Weibull 
distribution, with 
$$f(y,\theta,\gamma)=\exp\{-(\theta y)^\gamma\}
	\,\gamma(\theta y)^{\gamma-1}\theta, \quad y>0, \eqno(1.1)$$
then we should conceivably use 
$\hatt\mu_{\rm wide}=(\log 2)^{1/\hatt\gamma}/\hatt\theta$,
using ML estimators $\hatt\theta$, $\hatt\gamma$ in the wider
Weibull model. But {\it if} the simple model is right, i.e.~$\gamma=1$,
then $\hatt\mu_{\rm narr}$ is better, in terms (for example) of
mean squared error. By sheer continuity it should be
better also for $\gamma$'s close to 1. How much must $\gamma$ 
differ from 1 in order for $\hatt\mu_{\rm wide}$ to become better?
And what with similar questions for other typical 
parametric departures from exponentiality, like the gamma family?

\subsection
{\csc Example B.}
The most popular modelling of 
data $Y_1,\ldots,Y_n$ is to postulate normality, 
i.e.~assuming $f(y)=\phi((y-\xi)/\sigma)/\sigma$
for suitable parameters $\xi$ and $\sigma$. 
In many situations the normal density is too light-tailed
to constitute a serious description, however. 
A remedy then is to use
$$f(y,\xi,\sigma,m)=g_m\Bigl({y-\xi\over \sigma}\Bigr){1\over \sigma},$$
where $g_m(t)$ is the $t$-density with $m$ degrees of freedom. 
The narrower normal model corresponds to $m=\infty$,
and it is naturally felt that for large $m$
the discrepancy between normality and $t$-ness shouldn't matter.
Suppose for example that the parameter to be estimated is ${\rm sd}$,
the standard deviation for $Y_i$'s. 
For how large $m$ is it the case that the narrow-model
estimator $\hatt{\rm sd}_{\rm narr}$, 	
which happens to be the ordinary 
empirical standard deviation, is better than the
more laborious 
$$\hatt{\rm sd}_{\rm wide}=\sqrt{{\hatt m\over \hatt m-2}}\,\hatt\sigma,$$
computed from ML estimates $\hatt\xi$, $\hatt\sigma$, $\hatt m$
in the three-parameter model? 
What with other parameters to estimate than the standard deviation?

\subsection
{\csc Example C.}
Consider a regression situation with $n$ pairs $(x_i,Y_i)$.
The classical model says $Y_i\sim \normal\{\alpha+\beta x_i,\sigma^2\}$
for appropriate parameters $\alpha$, $\beta$, $\sigma$,
and encourages for example 
$\hatt\mu_{\rm narr}=\hatt\alpha_{\rm narr}+\hatt\beta_{\rm narr}x$
as the estimator for the median (or mean value) of the distribution
of $Y$ for a given $x$ value. Suppose however that the regression
curve could be moderately quadratic,
say $Y_i\sim \normal\{\alpha+\beta x_i+\gamma(x_i-\bar x)^2,\sigma^2\}$
for a moderate $\gamma$. How much must $\gamma$ differ from zero 
in order for 
$$\hatt\mu_{\rm wide}=\hatt\alpha+\hatt\beta x
	+\hatt\gamma(x-\bar x)^2,$$
with regression parameters now evaluated in the wider model, 
to perform better? The same questions could be
asked for other parameters, like comparing 
$\hatt x_{0,\rm narr}$ with $\hatt x_{0,\rm wide}$,
the narrow-model based and the wide-model based 
estimators of the point $x_0$ at which the regression 
curve crosses a certain level. And finally similar questions
could be discussed in the framework of am omitted covariate. 

\subsection
{\csc Example D.} 
In some situations a more interesting departure from 
standard regression lies in variance heterogeneity. 
This could for example suggest using 
$Y_i\sim \normal\{\alpha+\beta x_i,\sigma^2(1+\gamma x_i)\}$,
where $\gamma$ is zero under classical regression.
For what range of $\gamma$ values are standard methods, 
all derived under the $\gamma=0$ hypothesis, still better 
than four-parameter-model analysis?   

\subsection
{\csc Example E.}
Let us also include another type of model uncertainty,
that of misspecification due to using an incorrect transformation.
%The Box--Cox method is one method for 
%obtaining $Y_i'=T(Y_i)$ from original $Y_i$'s, with
%the property that the new $Y_i'$ is approximately 
%$N(\alpha+\beta x_i,\sigma^2)$. Since there is an estimated 
%parameter $\lambda$ present in the transformation  
%the regression analysis that follows is in principle affected
%by this, but this extra sampling variability is usually ignored;
%see Box--Cox--Bickel--Doksum ... The methods of this paper
%can be used to study this effect.
%We choose here to study the transformation problem for 
%another method, partly because there are some awkward points
%regarding parameter-dependent sample space for $Y_i'$.
%The transformation model invented here is as well-intentioned 
%as the Box--Cox method, but avoids this particular pitfall.
The transformation model invented here has some of the 
intentions of the Box--Cox power transformation scheme, 
but avoids some of its pitfalls.
It postulates that 
$$h_\lambda\bigl((Y_i-\alpha-\beta x_i)/\sigma\bigr)\sim \normal\{0,1\},
	\quad {\rm where} \quad
	h_\lambda(z)=\Phi^{-1}\{\Phi(z)^\lambda\}, \eqno(1.2)$$
for appropriate values of
$(\alpha,\beta,\sigma,\lambda)$; $\lambda=\lambda_0=1$
brings us back to classics. 
% --- can I write this ? ---
Let us briefly discuss this model and its use before 
we concentrate on the local misspecification part.  
It can be written 
$Y_i=\alpha+\beta x_i+\sigma Z_i$, where $h_\lambda(Z_i)$ follows 
the standard normal distribution for suitable 
transformation parameter. Varying $\lambda$ gives a fair range
of transformations, and in particular includes the possibility
of having skewed error distributions. 
The four parameters can be estimated from the data.
The notation is possibly deceiving in that it invites
one to think in terms of `$\alpha+\beta x_i$ plus noise with level $\sigma$'.
This is not quite the case since $Z_i$ has a skewed distribution 
with mean and median different from zero. 
It is advisable to reparameterise, 
after having found a suitable $\lambda$ from data, 
to the familiar structure + noise form. 
One possibility is 
$Y_i=\{\alpha+\sigma e(\lambda)\}+\beta x_i+\sigma v(\lambda)Z_i^0$,
in which $e(\lambda)$ and $v(\lambda)$ are mean value and standard
deviation of $Z_i$, under the $h_\lambda(Z_i)\sim \normal\{0,1\}$ model,
and where $Z_i^0$ now has mean zero and standard deviation 1.
Another possibility is 
$$\eqalign{
Y_i&=\{\alpha+\sigma\Phi^{-1}(0.50^{1/\lambda})\}+\beta x_i
	+\sigma\{\Phi^{-1}(0.75^{1/\lambda})
		-\Phi^{-1}(0.25^{1/\lambda})\}Z_i' \cr
   &=\alpha'+\beta x_i+\sigma'Z_i', \cr} \eqno(1.3)$$ 
the point being that $Z_i'$ has median zero and interquartile range 1.
Our technical point is that (1.2) is a useful generalisation of  
classical regression to situations with skewed errors, and that
parameter estimation is perhaps best carried out using ML machinery 
on (1.2); and our statistical point is that (1.3) better conveys 
the structure and the noise in the data, and should be used 
post estimation. 
 
The present concern is how robust standard methods, 
which presume $\lambda=1$, are against misspecification of that parameter.  
Should one use  
$$\hatt\mu_{\rm wide}(x)=\hatt\alpha_{\rm wide}+\hatt\beta_{\rm wide}x
	+\hatt\sigma_{\rm wide}
	 \Phi^{-1}(0.50^{1/\hatt\lambda_{\rm wide}}) \eqno(1.4)$$
to estimate the median of $Y$ for given $x$, or will the 
%less ambitious and 
effortlessly obtainable 
$\hatt\mu_{\rm narr}(x)=\hatt\alpha_{\rm narr}+\hatt\beta_{\rm narr}x$ 
suffice? 

\subsection
{\csc Example F.} 
% 	Nils: if the next goes out then use ``Our final example...'' here.
Next consider logistic regression, 
in which pairs $(x_i,Y_i)$ are observed of the type
$Y_i\midd x_i\sim{\rm Bin}\{1,p(x_i)\}$, with 
$p(x)=\exp(\alpha+\beta x)/\{1+\exp(\alpha+\beta x)\}$ being 
the standard model. Again we can ask whether 
standard methods based on $(\hatt\alpha_{\rm narr},\hatt\beta_{\rm narr})$,
for example for estimating the true $p(x)$ at a given $x$, 
or for estimating the cut-off point at which $p(x)$ exceeds $\half$,
become seriously inferior under moderate misspecifications.
One natural type of departure is modelled by adding  
a quadratic term $\gamma(x_i-\bar x)^2$ to the linear term;
another is 
$$p(x)=p(x,\alpha,\beta,\eta)=\Bigl\{
	{\exp(\alpha+\beta x)\over 1+\exp(\alpha+\beta x)}\Bigr\}^\eta,
							\eqno(1.5)$$
where it is of interest to vary $\eta$ around $\eta_0=1$.

\subsection
{\csc Example G.}
Our final example is the two-sample model with variances that 
may or may not be equal. So $X_1,\ldots,X_m$ are $\normal\{\xi_1,\sigma_1^2\}$
and $Y_1,\ldots,Y_n$ are $\normal\{\xi_2,\sigma_2^2\}$, all of them are
independent, and the narrow model specifies that $\sigma_1=\sigma_2$.
Under this assumption it is easy to put up estimators, confidence
intervals etc.~for parameters related to the difference between
the $X$-distribution and the $Y$-distribution, like 
the Mahalanobis distance $\Delta=|\xi_2-\xi_1|/\sigma$. 
More awkward methods are needed when $\sigma_2\not=\sigma_1$,
cf.~the Behrens--Fisher problem. The in some sense natural
generalisation of the Mahalanobis distance is for example
$$\Delta=(\nu^2+\omega^2)^{1/2}, 
	\quad {\rm where\ }\nu^2=(\xi_2-\xi_1)^2/\sigma^2,\,
	      \omega^2=4\log{\sigma^2\over \sigma_1\sigma_2},\,
	      \sigma^2=(\sigma_1^2+\sigma_2^2)/2,$$ 
see Hjort (1986a, Ch.~10). How resistant is the simple 
$\hatt \Delta_{\rm narr}=|\bar Y-\bar X|/\hatt\sigma_{\rm narr}$
to differences in $\sigma_1$, $\sigma_2$? When is it necessary 
to use the much more complicated $\hatt \Delta_{\rm wide}$? 

\subsection
Let us summarise the common characteristics of these situations.
There is a narrow and usually simple parametric model 
%	$f(y,\theta)$ 
which can be fitted to the data, 
but there is a potential misspecification,
which can be ameliorated by its encapsulation in a wider model 
%	$f(y,\theta,\gamma)$ 
with one additional parameter. 
Estimating a parameter assuming correctness of the narrow model 
involves modelling bias, 
but doing it in the wider model could mean larger 
sampling variability. Thus the choice of method 
becomes a statistical balancing act with perhaps deliberate bias 
against variance. 
% Note that the parameter to be estimated is 
% meant to be an identifiable function of the wide model itself ... 
	
The examples above span a reasonable range 
of heavily used `narrow' models 
along with indications of rather typical kinds of deviances from them.
Many standard textbook methods for parametric inference 
are derived under the conditions of such narrow models. 
Our main result, derived in Section 3,
is a surprisingly sharp and general large-sample criterion 
for how much misspecification a given narrow model can
tolerate. This criterion is applied to Examples A--G in
Section 7. It is relatively easy to compute, in that it only
involves the familiar Fisher information matrix, for the wide
model, but evaluated under narrow model conditions. 
A particular
% pleasing nor not pleasing, c'est la question 
facet of our tolerance criterion is that it does not depend upon 
the particular parameter estimand at all.

In addition to quantifying the degree of robustness of 
standard methods there are also 
pragmatic reasons for the present investigation.
Statistical analysis will in practice still be carried out
using narrow model based methods in the majority of cases, 
for reasons of ignorance, simplicity, na\"\i vit\'e and boldness; 
using wide model methods will very often be much more laborious,
and only experts will use them anyhow. 
Thus it is of interest to quantify the consequences of ignorance,
and it would be nice to obtain permission to go on 
doing analysis as if the simple model were true.
Such a partial permission is in fact given here.
% under moderate but explicit constraints on the amount 
% of misspecification.  
% see section 7 for examples. 
% and sections 4 and 5 for discussion
% and the unavoidable caveat. 
%	Nils: Caveat: only about precision of estimates,
%	not preciseness or correctness of following analysis.
The results of this paper can be interpreted as 
saying that `ignorance is (sometimes) strength)'; 
mild departures from the narrow model do not really matter, 
and more ambitious methods could perform worse. 
% when it comes to precision of estimators. 
In the examples of Section 7 quite explicit limits are given 
for the degree of misspecification that is tolerable. 
This upper limit is in most cases 
dependent upon parameters of the model, and should be estimated
by the conscientious statistician in situations where 
departures of the type described are suspected. 
The compromise estimators that we advocate in Section 5 
utilise this departure estimate.
% So one should by all means carry out estimation of the additional
% parameter, even if it turns out that 
% it won't be needed in the final analysis.

Several tangential topics are taken up in Section 4.
These include measures of distance
from null model to the least tolerable misspecification;
comparison with the model selection criteria of Akaike and Schwarz; 
simulation based evaluation of our criterion;
discussion of the concept of a robust model; 
dangerous versus noncritical departures from a model;
and interpretation of confidence intervals under misspecification.
Deviances from a model in more than one direction is briefly discussed 
at the end of Section 5. 
%	Nils: alles in Ordnung? 

There is also room for improvement over the narrow and wide methods. 
In Section 5 some new estimators are proposed that are designed 
to work well both under narrow and wide circumstances. 
A broad comparison of the various compromise estimators is made,
in a large-sample framework of moderately misspecified parametric models. 
A connection to Bayesian robustness is also made. 
We are able to make a quite general and drastic reduction: 
The performance of a large class of competing estimators  
can be studied in a much simpler and very classical context, 
that of estimating $a$ in a $\normal\{a,1\}$ situation with one observation! 
Here the narrow model corresponds to $a=0$. 
This provides fresh motivation for studying 
$a$-estimators that in various ways take into account that
values of $a$ in the vicinity of zero are perhaps 
more likely or perhaps more important. 
Such a study is reported on in Sections 5 and 6. 
%	Nils: More? It is instructive to see them naked.

The traditional robustness literature is mostly 
concerned with construction of methods that perform
well over a `nonparametric neighbourhood' around some basic model. 
The present work is different in that it envisages 
specific, parametric alternatives to the basic model.   
There is a literature on parametric robustness,
perhaps chiefly concerned with studying behaviour of 
standard methods and modified standard methods 
under natural violations of the basic model.
Only rarely have comparisons been made between `narrow' and `wide'
methods, however. 
Some papers have calculated and commented on 
the increased estimation noise for a narrow model parameter 
when passing to a wider model, like comparing the variances of 
$\hatt\theta_{\rm narr}$ and $\hatt\theta_{\rm wide}$ in Example A.
This is beside the point, partly confusing, and not very interesting, 
since what matters is studying 
`real' parameters which are meaningful functions of the full model, as 
the median $\mu=\mu(f)=\mu(\theta,\gamma)=(\log2)^{1/\gamma}/\theta$ 
in Example A, the standard deviation
${\rm sd}={\rm sd}(f)={\rm st}(\xi,\sigma,m)=\{m/(m-2)\}^{1/2}\sigma$ in
Example B, etcetera. 
Bickel (1984) is on the other hand clear about this issue, 
and is concerned with several problems 
that resemble those considered here. 
He does not compare narrow and wide methods, 
and does not study tolerance distances, 
but works directly with certain minimax strategies,
in a framework of nested linear normal models; 
see also 5G below. 
The paper by Berger (1982) on Bayesian robustness
also turns out to be related to some of these questions.
See Bickel's comments on Berger and 5E, 5F, 5G below. 
%	--- Be careful, Nils? --- 

\bigskip
{\bf 2. Large-sample framework for the problem.}
We shall start our investigation in the i.i.d.~framework.
Suppose $Y_1,\ldots,Y_n$ come from some
common density $f$, and represent the wide model
as $f(y)=f(y,\theta,\gamma)$, where $\gamma=\gamma_0$ 
corresponds to the narrow model, say 
$f(y,\theta)=f(y,\theta,\gamma_0)$. 
We assume that $\theta=(\theta_1,\ldots,\theta_p)'$ lies
in some open region in Euclidean $p$-space, that 
$\gamma$ lies in some open interval containing $\gamma_0$, 
and that the wide model is `smooth';
for definiteness we postulate that the regularity conditions
put forward in Lehmann's (1983) chapter 6.4 are in force.   
We are to study 
behaviour of estimators when $\gamma$ deviates from $\gamma_0$. 
The parameter to be estimated is some $\mu=\mu(f)$,
which we write as $\mu(\theta,\gamma)$ since the wider model 
is assumed to be an adequate description of reality. 
We concentrate on ML procedures, and write $\hatt\theta_{\rm narr}$ for
the estimator of $\theta$ in the narrow model and 
$(\hatt\theta,\hatt\gamma)$ for the estimators in the wide model.
The two major entries in the competition are 
$$\hatt\mu_{\rm narr}=\mu(\hatt\theta_{\rm narr},\gamma_0)
	\quad {\rm and} \quad
	\hatt\mu_{\rm wide}=\mu(\hatt\theta,\hatt\gamma) \eqno(2.1)$$
(but see Section 5 for other estimators).  

These could be compared in an 
asymptotic framework in which $Y_i$'s come from some
fixed $f(y,\theta_0,\gamma)$, and $\gamma\not=\gamma_0$. 
%In the situation of Example A, the formulae given in section 7
%may be used to show that 
%$\sqrt{n}\{(\log 2)^{1/\hatt\gamma}/\hatt\theta-\mu\}$ 
%tends to a $\normal\{0,1.589(\log 2)^2/\theta_0^2\}$, 
%while 
%$$\sqrt{n}\Bigl({\log 2\over \hatt\theta_{\rm narr}}
%		-\mu\Bigr)
%%%%%%%		-{(\log 2)^{1/\gamma}\over \theta_0}\Bigr\}
%	=\sqrt{n}\Bigl({\log2\over \hatt\theta_{\rm narr}}
%		-{\log2\over \theta_0}\Bigr)
%	-\sqrt{n}\Bigl\{{(\log2)^{1/\gamma}\over \theta_0}
%		-{\log2\over \theta_0}\Bigr\}$$
%has two terms; the first tends to a $\normal\{0,1.109(\log2)^2/\theta_0^2\}$,
%reflecting smaller sampling variability, but second tends to plus of minus 
%infity, reflecting a bias that sooner or later will dominate completely.  
In this case $\sqrt{n}(\hatt\mu_{\rm wide}-\mu)$ has a limit
distribution, which can be derived from the proposition below. 
The situation is different for the narrow model procedure. Here 
$\sqrt{n}(\hatt\mu_{\rm narr}-\mu)$ 
can be represented as a sum of two terms.
The first is 
$\sqrt{n}\{\mu(\hatt\theta_{\rm narr},\gamma_0)-\mu(\theta_0,\gamma_0)\}$,
which has a limit distribution, with generally smaller 
variability than that of the wide model procedure, and the second is 	  
$-\sqrt{n}\{\mu(\theta_0,\gamma)-\mu(\theta_0,\gamma_0)\}$, 
which tends to plus or minus infinity, reflecting a bias 
that for very large $n$ will dominate completely. 
This merely goes to show that with very large sample sizes 
one is penalised for any bias and one should use 
the wide model. 
This result is not very informative, however, and 
suggests that a large-sample framework which uses
a local neighbourhood of $\gamma_0$ that shrinks 
when the sample size grows is much more adequate. 
Study therefore model $P_n$, the $n$'th model, under which 
$$Y_1,\ldots,Y_n{\rm \ are\ i.i.d.\ from\ } 
	f_n(y)=f(y,\theta_0,\gamma_0+\delta/\sqrt{n}), \eqno(2.2)$$ 
and where $\theta_0$ is fixed but arbitrary. 
In this framework we need limit distributions 
for the wide model estimators $(\hatt\theta,\hatt\gamma)$
and for the narrow model estimator $\hatt\theta_{\rm narr}$.

Consider 
$$\mtrix{U(y) \cr 
	   V(y) \cr}
=\mtrix{\dell\log f(y,\theta_0,\gamma_0)/\dell\theta \cr
	   \dell\log f(y,\theta_0,\gamma_0)/\dell\gamma \cr}, \eqno(2.3)$$
the score function for the wide model, but evaluated at the
null point $(\theta_0,\gamma_0)$.
The accompanying familiar $(p+1)\times(p+1)$ size information matrix is
$$J_{\rm wide}={\rm VAR}_0\mtrix{
	 \dell\log f(Y,\theta_0,\gamma_0)/\dell\theta \cr
	 \dell\log f(Y,\theta_0,\gamma_0)/\dell\gamma \cr}
	=\mtrix{J_{11} &J_{12} \cr
		  J_{21} &J_{22} \cr}. $$
Note that the $p\times p$ size $J_{11}$ is simply the information matrix 
of the narrow model, evaluated at $\theta_0$,
and that the scalar $J_{22}$ is the variance of 
$V(Y_i)$, also computed under the narrow model.

\smallskip
{\csc Proposition.} {{\sl   
Under the sequence of models $P_n$ of (2.2), 
as $n$ tends to infinity, we have
$$\mtrix{\sqrt{n}(\hatt\theta-\theta_0) \cr 
	    \sqrt{n}(\hatt\gamma-\gamma_0-\delta/\sqrt{n}) \cr}
	\rightarrow_d \normal_{p+1}\{0,J_{\rm wide}^{-1}\}, 
	\ {\sl or}\ 
   \mtrix{\sqrt{n}(\hatt\theta-\theta_0) \cr	
	    \sqrt{n}(\hatt\gamma-\gamma_0) \cr}
	\rightarrow_d \normal_{p+1}\{\mtrix{0 \cr \delta \cr},
					J_{\rm wide}^{-1}\};$$
$$\sqrt{n}\{\hatt\theta_{\rm narr}
	-(\theta_0+J_{11}^{-1}J_{12}\delta/\sqrt{n})\}
	\rightarrow_d\normal_p\{0,J_{11}^{-1}\}, 
	\ {\sl or} \
	\sqrt{n}(\hatt\theta_{\rm narr}-\theta_0)
	\rightarrow_d\normal_p\{J_{11}^{-1}J_{12}\delta,J_{11}^{-1}\}.$$
}}
%	-----------------------
%	What is really required? 
%We use the following slightly extended version of the Lindeberg theorem:
%Let $Y_{n,i}$ i.i.d.~with $\mu_n$ and $\tau_n^2$.
%Suppose $\sqrt{n}\mu_n\rightarrow\mu$ and $\tau_n^2\rightarrow\tau^2$.
%Assume that 
%$$EY_{n,i}^2I\{|Y_{n,i}|\ge\eps\sqrt{n}\}=\int y^2f_n(y)\,\d y\rightarrow0.$$
%Then $\sqrt{n}\bar Y_n\rightarrow_d\normal\{\mu,\tau^2\}$.
%In our case: 
%$\int U_a(x)^2f(x,\theta_0,\gamma)\,\d x$ continuous at $gamma=\gamma_0$,
%$\int V(x)^2f(x,\theta_0,\gamma)\,\d x$ continuous at $gamma=\gamma_0$:
%This implies Lindeberg condition and $\tau_n^2\rightarrow\tau^2$. 
%Also, to ensure rest:
%$\{\dell/\dell\gamma\}_{\gamma=\gamma_0}\int U(x)f(x,\theta_0,\gamma)\,\d x$
%should be $\int UV f_\theta\,\d x=J_{12}$, and similarly
%$\{\dell/\dell\gamma\}_{\gamma=\gamma_0}\int V(x)f(x,\theta_0,\gamma)\,\d x$
%should be $\int V^2 f_\theta\,\d x=J_{22}$. 
%	Need to sort out whether Lehmann's conditions suffice. 
%	------------------------------------------------------

{\csc Proof:} 
Consider $\hatt\theta_{\rm narr}$ first. 
The familiar Taylor expansion arguments that lead to the 
classical $\sqrt{n}(\hatt\theta_{\rm narr}-\theta_0)\rightarrow_d
\normal\{0,J_{11}^{-1}\}$ under the null model $f(x,\theta_0,\gamma_0)$ 
can be used in the present $\gamma_0+\delta/\sqrt{n}$ case as well. For 
$$\sum_{i=1}^n{\dell\over \dell\theta}
	\log f(Y_i,\hatt\theta_{\rm narr},\gamma_0)
	=\sum_{i=1}^nU(Y_i)+I_n(\tilda\theta_n)
		(\hatt\theta_{\rm narr}-\theta_0)=0,$$
in which $I_n(\theta)=\sum_{i=1}^n\dell^2
\log f(Y_i,\theta,\gamma_0)/\dell\theta\dell\theta'$ 
and $\tilda\theta_n$ lies somewhere between $\theta_0$ 
and $\hatt\theta_{\rm narr}$. Under the conditions stated 
$\hatt\theta_{\rm narr}\rightarrow_p\theta_0$,
under $P_n$, using necessary but moderate 
% slight 
variations of the arguments used in
Lehmann's (1983) chapter 6.4 and 6.8, 
and $-I_n(\theta_0)/n$ as well as $-I_n(\tilda\theta_n)/n$ 
tend in probability, under $P_n$, to $J_{11}$. All this leads to 
$$\sqrt{n}(\hatt\theta_{\rm narr}-\theta_0)\doteq_d
	\{\hbox{$-{1\over n}$}I_n(\theta_0)\}^{-1}\sqrt{n}\bar U_n
	\doteq_d J_{11}^{-1}\sqrt{n}\bar U_n, \eqno(2.4)$$
where $A_n\doteq_dB_n$ means that $A_n-B_n$ tends to zero in probability,
and $\bar U_n$ is the average of the $n$ first $U(Y_i)$'s. 
The triangular version of the Lindeberg theorem shows that
%	But $-I_n(\theta_0)/n$ tends in probability, under $P_n$, to $J_{11}$,
$\sqrt{n}\bar U_n$ tends in distribution, under $P_n$, 
to $\normal_p\{J_{12}\delta,J_{11}\}$. This is because
$$\eqalign{
\E_{P_n}U(Y_i)&=\int f(y,\theta_0,\gamma_0+\delta/\sqrt{n})U(y)\,\d y \cr
	&\doteq\int f(y,\theta_0,\gamma_0)
		\{1+V(y)\delta/\sqrt{n}\}U(y)\,\d y 
	 	=J_{12}\delta/\sqrt{n}, \cr}$$
and similar calculations show that $U(Y_i)U(Y_i)'$ has expected value 
$J_{11}+O(\delta/\sqrt{n})$, under $P_n$. 
%Also, under the conditions
%stated $\hatt\theta\rightarrow_p\theta_0$, using slight variations
%of the arguments used in Lehmann's (1983) chapter 6.4 and 6.8, 
%and $-I_n(\theta_0)/n$ tends in probability, under $P_n$, to $J_{11}$. 
This proves the `narrow' part of the proposition.

Similar reasoning takes care of the `wide' part too. One finds
$$\mtrix{\sqrt{n}(\hatt\theta-\theta_0) \cr
	   \sqrt{n}(\hatt\gamma-\gamma_0) \cr}
	\doteq_d J_{\rm wide}^{-1}
	\mtrix{\sqrt{n}\bar U_n \cr
		 \sqrt{n}\bar V_n \cr}
	\rightarrow_d
	J_{\rm wide}^{-1}\normal_{p+1}\{
	\mtrix{J_{12}\delta \cr J_{22}\delta \cr},J_{\rm wide}\},\eqno(2.5)$$
which is equivalent to the wide part statement. 
More details with reference to a precise set of regularity conditions 
are in K\aa resen (1992). \square

\smallskip
{\csc Remark.} 
Let us for a moment consider more general departures from
the $f(y,\theta)$ model. 
Assume only that data $Y_i$ come from a fixed $f$.
Then $\hatt\theta_{\rm narr}$ is consistent
for the particular `least false' or `most appropriate' 
parameter value $\theta_{\rm l.f.}=\theta(f)$ that 
minimises the Kullback--Leibler distance 
$d[f\colon f(.,\theta)]=\int f(y)\log\{f(y)/f(y,\theta)\}\,\d y$.
One can also show that $\sqrt{n}(\hatt\theta_{\rm narr}-\theta_{\rm l.f.})$
tends in distribution to $\normal_p\{0,J(f)^{-1}K(f)J(f)^{-1}\}$,
in which 
$$J(f)=-\E_f{\dell^2\log f(Y,\theta_{\rm l.f.})\over \dell\theta\dell\theta},
	\quad {\rm and} \quad
	K(f)={\rm VAR}_f{\dell\log f(Y,\theta_{\rm l.f.})
		\over \dell\theta}. \eqno(2.6)$$
This is for example made clear in Hjort (1986a, Ch.~3). --- Let us
%	Nils: should I include other references? Like Hjort (1988)?  
apply this to the local misspecification situation, that is,
insert $f(y)=f(y,\theta_0,\gamma)$, where $\gamma$ is close to $\gamma_0$. 
Then judicious Taylor expansion arguments show that
$$\theta_{\rm l.f.}=\theta_0+J_{11}^{-1}J_{12}(\gamma-\gamma_0)
	+O((\gamma-\gamma_0)^2), \quad 
	J(f)^{-1}K(f)J(f)^{-1}=J_{11}^{-1}+O(\gamma-\gamma_0).$$
Using this, for the local $\gamma=\gamma_0+\delta/\sqrt{n}$, 
can be used to prove the `narrow part' of the proposition again.
% 	* Also: regularity of estimator sequences. *
%	--- Nils: interpretation of parameters change --- 
%	--- Nils: could expand on the variance formula ---
Note that the notion and interpretation of a 
best fitting parameter changes when the model changes, 
and that the results about $\theta_{\rm l.f.}$ 
quantify this in a precise way.

\bigskip
{\bf 3. Solution.}
% Suppose $\mu$ is a parameter to be estimated. It is a function of 
% the true probability distribution, so let us write 
% $\mu=\mu(\theta,\gamma)$, and assume that it is smooth in the
% sense of having continuous partial derivatives 
% w.r.t.~$\theta$ and $\gamma$.   
In the large-sample framework of the previous section we are to
compare two estimators: The `safe' 
$\hatt\mu_{\rm wide}=\mu(\hatt\theta,\hatt\gamma)$ based on
ML estimation in the wide model, and the `risky'
$\hatt\mu_{\rm narr}=\mu(\hatt\theta_{\rm narr},\gamma_0)$.
The true parameter is 
$\mu_{\rm true}=\mu(\theta_0,\gamma_0+\delta/\sqrt{n})$
under $P_n$, the $n$'th model. 
Our comparison criterion is the limit of sample size times 
mean squared error; see 5H for a technical comment and for 
other possibilities.

First consider the safe estimator. By the delta method
of linearisation we find
$$\eqalign{
\sqrt{n}\{&\mu(\hatt\theta,\hatt\gamma)
	-\mu(\theta_0,\gamma_0+\delta/\sqrt{n})\} \cr
	&\doteq_d (\delltheta)'\sqrt{n}(\hatt\theta-\theta_0)
	 +\{(\dellgamma)+O(1/\sqrt{n})\}
	  \sqrt{n}(\hatt\gamma-(\gamma_0+\delta/\sqrt{n})) 
	  \rightarrow_d \normal\{0,\tau^2\}, \cr}$$
where 
$$\tau^2=\mtrix{\delltheta \cr
		  \dellgamma \cr}'J_{\rm wide}^{-1}
		\mtrix{\delltheta \cr
		 	 \dellgamma \cr}. \eqno(3.1)$$
The partial derivatives are computed at the null point $(\theta_0,\gamma_0)$. 
Similarly, for the risky estimator,
$$\eqalign{
\sqrt{n}\{&\mu(\hatt\theta_{\rm narr},\gamma_0)
	-\mu(\theta_0,\gamma_0+\delta/\sqrt{n})\} \cr
	&=\sqrt{n}\{\mu(\hatt\theta_{\rm narr},\gamma_0)
		-\mu(\theta_0,\gamma_0)\}
	 -\sqrt{n}\{\mu(\theta_0,\gamma_0+\delta/\sqrt{n})
		-\mu(\theta_0,\gamma_0)\} \cr
	&\doteq_d(\delltheta)'\sqrt{n}(\hatt\theta_{\rm narr}-\theta_0)
		-\sqrt{n}\dellgamma\delta/\sqrt{n} 
	 \rightarrow_d\normal\{b\delta,\tau_0^2\}, \cr}$$
in which 
$$b=J_{21}J_{11}^{-1}(\delltheta)-\dellgamma
	\quad {\rm and} \quad
	\tau_0^2=(\delltheta)'J_{11}^{-1}(\delltheta). \eqno(3.2)$$
By evaluating the mean value of the square of the limit distributions  
we have that $n$ times the asymptotic mean squared error 
of $\hatt\mu_{\rm wide}$ becomes $\tau^2$, 
while the corresponding quantity for $\hatt\mu_{\rm narr}$
becomes $b^2\delta^2+\tau_0^2$.
      
We are now in a position to find out when 
the risky estimator is better than the safe one, 
simply by algebraically solving the inequality 
$b^2\delta^2+\tau_0^2\le \tau^2$ w.r.t.~$\delta$. 
Start out writing 
$$J_{\rm wide}^{-1}=\mtrix{J^{11} &J^{12} \cr
			     J^{21} &J^{22} \cr},$$
where a prominent r\^ole is designated for 
$$J^{22}=\kappa^2=(J_{22}-J_{21}J_{11}^{-1}J_{12})^{-1} \eqno(3.3)$$
in what follows, and 
$J^{12}=-J_{11}^{-1}J_{12}\kappa^2$, 
$J^{11}=J_{11}^{-1}+J_{11}^{-1}J_{12}J_{21}J_{11}^{-1}\kappa^2$. 
This leads to the simplification
$$\eqalign{
\tau^2&=(\delltheta)'J_{11}^{-1}(\delltheta)
	+(\delltheta)'J_{11}^{-1}J_{12}J_{12}'J_{11}^{-1}(\delltheta)\kappa^2
	-2(\delltheta)'J_{11}^{-1}J_{12}(\dellgamma)\kappa^2
	+(\dellgamma)^2\kappa^2 \cr
	&=\tau_0^2+b^2\kappa^2. \cr}$$
We have reached the following. 

\smallskip
{\csc Result.} {{\sl 
(i) The case where $b=0$ is rather trivial; 
this typically corresponds to 
asymptotic independence between $\hatt\theta$ and $\hatt\gamma$ 
under the null model, and a parameter $\mu$ functionally independent 
of $\gamma$. In this case $\hatt\mu_{\rm wide}$ and $\hatt\mu_{\rm narr}$ 
are asymptotically equivalent, regardless of $\delta$. 
(ii) In the more interesting case
$b\not=0$, the narrow model based estimator is better than or as good
as the wider model based estimator 
$${\sl if\ and\ only\ if}\quad \delta^2\le\kappa^2,
\quad{\sl or\ }|\delta|\le\kappa,
\quad{\sl or\ }|\gamma-\gamma_0|\le\kappa/\sqrt{n}. \eqno(3.4)$$
\smallskip}}

{\sl Extension to regression models.} 
To solve the problems raised in the regression type examples of 
the introduction we also need the analogous result in the more 
general situation of independent observations with covariates.
This can be done in a fairly straightforward fashion.  
Examples C--F of Sections 1 and 7 lead us naturally to 
the following general framework. 
Suppose $(x_i,Y_i)$ are independent pairs, where $Y_i$
has density $f(y_i,\sigma,\beta,\gamma\midd x_i)$ for given $x_i$-value,
carrying some scale parameter $\sigma$ (but not necessarily),
a vector $\beta=(\beta_1,\ldots,\beta_p)'$ 
of ordinary regression parameters, 
plus some interesting one-dimensional
extra parameter $\gamma$ that signals departure from the underlying 
classical model, which corresponds to some appropriate $\gamma=\gamma_0$.

Under mild regularity conditions the main result  
above continues to be true for regression models,
with $\kappa^2$ defined as in (3.3), but with a somewhat more
cumbersome $J_{\rm wide}$ matrix than before. 
The correct definition is now 
$$J_{\rm wide}=\lim_{n\rightarrow\infty} J_{n,\rm wide}
	=\lim_{n\rightarrow\infty}
	{1\over n}\sum_{i=1}^n{\rm VAR}_0
  \mtrix{\dell\log f(Y_i,\sigma_0,\beta_0,\gamma_0\midd x_i)/\dell\sigma \cr
	   \dell\log f(Y_i,\sigma_0,\beta_0,\gamma_0\midd x_i)/\dell\beta \cr
	   \dell\log f(Y_i,\sigma_0,\beta_0,\gamma_0\midd x_i)/\dell\gamma \cr},
					\eqno(3.5)$$
where the variance matrices are computed at the null model,
under $(\sigma_0,\beta_0,\gamma_0)$. 
The necessary regularity conditions can be put up in various forms.
These would be Lindebergian to secure normal limits 
and must in particular imply convergence of $J_{n,\rm wide}$; 
this usually follows if it is assumed that
the collection of $x_i$'s come from some distribution in the 
design space. See K\aa resen (1992) for a detailed argument.
In practice one would typically use $J_{n,\rm wide}$
to compute $\kappa^2$. 
Examples are given in Section 7.

\bigskip
{\bf 4. Discussion.}

\subsection
{\sl 4A. Simplicity.} 
It is remarkable that the criterion (3.4) does not depend 
on the particularities of the specific parameter $\mu(\theta,\gamma)$ 
at all. Thus, in the situation of Example A in the introduction,
calculations in Section 7 show that 
$|\gamma-1|\le 1.245/\sqrt{n}$ guarantees
that being simple-minded, assuming exponentiality, 
works better than being ambitious, using a gamma-family, 
for {\it every} smooth parameter $\mu(\theta,\gamma)$. 
(This is different in a situation with more a multi-dimensional
departure from the model, see 5I.)

Our criterion $\delta^2\le\kappa^2$ can be evaluated and assessed 
just from knowledge of $J_{\rm wide}$, the information matrix
of the full model, but computed at the narrow model only. 
This is fortunate, as the general $p+1$ parameter matrix will
be very hard to compute in many applications, but 
will be simpler and manageable at the null model. 
This is demonstrated in Section 7.  
Observe that the $|\delta|\le\kappa$ criterion can be thought of in terms 
of the limiting variance for $\hatt\gamma$, at the null model,
since $\sqrt{n}(\hatt\gamma-\gamma_0)$ tends to $\normal\{0,\kappa^2\}$ then. 

\subsection
{\sl 4B. How far away is the border line?} 
We have shown that the simple $\theta$ parameter model can tolerate
up to $\gamma_0+\kappa/\sqrt{n}$ deviation from $\gamma_0$ in the
encapsulating $(\theta,\gamma)$ model. 
How far is the border line $\delta=\kappa$ from the narrow model? 
One way of answering this is in terms of the probability of 
actually detecting that the narrow model is wrong. 
The natural 5\% level test for the 
correctness of the narrow model, against the alternative hypothesis that
the additional $\gamma$ parameter must be included, is to reject when
$Z_n^2=n(\hatt\gamma-\gamma_0)^2/\hatt\kappa^2$ exceeds $1.96^2$, since  
$Z_n^2$ has a limiting $\chi^2_1$ distribution under the narrow model. 
Here $\hatt\kappa$ is any consistent estimator of $\kappa$,
or simply equal to the known value in such cases.
%	Nils: (see the examples of section 7). 
The probability that this test 
detects that $\gamma$ is not equal to $\gamma_0$, 
when it in fact is equal to $\gamma_0+\delta/\sqrt{n}$, converges to 
$${\rm power}(\delta)
	=\Pr\{\chi^2_1(\delta^2/\kappa^2)>1.96^2\},\eqno(4.1)$$
featuring the non-central chi squared with 1 degree of freedom and 
eccentricity parameter $\delta^2/\kappa^2$.
This is a consequence of the proposition proved in Section 2.  
In particular the approximate power at the border case is equal to 17.0\%.
We can therefore restate the basic result as follows:
Provided the true model deviates so modestly from the narrow model
that the probability of detecting it is 17.0\% or less with the
natural 5\% level test, then the risky estimator is better than  
the safe estimator. Corresponding other figures 
for (level, power) are, for illustration, 
(0.01, 0.057), 
% (0.05, 0.170),
(0.10, 0.264),
(0.20, 0.400), 
(0.29, 0.500).
%	--- Nils: satisfied? 

\subsection
{\sl 4C. Other distance measures.} 
Let us present a couple of further measures of the distance
from null model to border line misspecification.
(i) 
The Kullback--Leibler distance 
$d[f(.,\theta_0,\gamma_0)\colon 
	f(.,\theta_0,\allowbreak \gamma_0+\delta/\sqrt{n})]$
can by Taylor expansion arguments be shown to be
equal to $\half \delta^2J_{22}/n$ plus smaller terms,
and in the border case the distance becomes 
$\kappa^2J_{22}/2n$. 
(ii) 
Next consider the so-called statistical distance or $L_1$-distance
between the two neighbouring distributions. It is
$$\int|f(y,\theta_0,\gamma_0+\delta/\sqrt{n})
	-f(y,\theta_0,\gamma_0)|\,\d y\doteq 
	 {\delta\over \sqrt{n}}\int |V(y)|\,f(y,\theta_0,\gamma_0)\,\d y.$$
This distance has a direct probabilistical interpretation.
In Example A, for example, the $L_1$-distance from exponentiality 
to the least tolerable Weibull, becomes about $0.923/\sqrt{n}$. 
%	Weibull: $\int_0^\infty |1+\log x-x\log x|\,exp(-x)\,\d x=0.923$.
%	Gamma: $\int_0^\infty |\log x+0.5772|\,exp(-x)\,\d x=0.983$.
(iii) Finally consider weighted $L_2$-distance 
$\int (f-f_0)^2/f_0\,\d y$. An approximation is seen to be
$\delta^2J_{22}/n$, and the least tolerable distance is
$\kappa^2J_{22}/n$. --- Note that these three distance 
measures are transformation invariant. See also 4F. 

\subsection
{\sl 4D. Comparison with Akaike's Information Criterion.} 
The misspecification problem is related to that of choosing a model.
One general method for doing this is to use the information criterion
of Akaike. In the present setting one is to compare
$${\rm AIC}_{\rm narr}=2\log L_{\rm max,narr}-2p 
	\quad {\rm with} \quad
	{\rm AIC}_{\rm wide}=2\log L_{\rm max,wide}-2(p+1),$$
featuring maximised log likelihoods under respectively 
the narrow model with $p$ parameters and the 
wide model with $p+1$ parameters. The method consists of 
choosing the model with largest observed AIC.
(Actually Akaike in his first work on this criterion used 
minus what we have taken the liberty of calling AIC here.
We prefer maximising likelihoods to minimising inverse likelihoods.)
The factor `2' is not important but is there since differences
between maximised nested log-likelihoods go to half chisquares 
under certain conditions, cf.~the deviance notion of generalised
linear models, and the calculations below. 

It is instructive to study AIC's behaviour in the framework of this
article. Using Taylor expansion, along with techniques and notation as in 
the proof of the proposition of Section 2, one finds 
$$\eqalign{
{\rm AIC}_{\rm narr}&\doteq_d 2\sum_{i=1}^n\log f(Y_i,\theta_0,\gamma_0)
	+n\bar U_n'J_{11}^{-1}\bar U_n-2p, \cr
{\rm AIC}_{\rm wide}&\doteq_d 2\sum_{i=1}^n\log f(Y_i,\theta_0,\gamma_0)
	+n\mtrix{\bar U_n \cr \bar V_n \cr}'
		J_{\rm wide}^{-1}\mtrix{\bar U_n \cr \bar V_n}-2(p+1). \cr}$$
Further algebraic calculations give 
$$\eqalign{ 
{\rm AIC}_{\rm wide}-{\rm AIC}_{\rm narr}
	&\doteq_d n\mtrix{\bar U_n \cr \bar V_n \cr}'
		J_{\rm wide}^{-1}\mtrix{\bar U_n \cr \bar V_n \cr}
		-n\bar U_n'J_{11}^{-1}\bar U_n-2 \cr
	&=n\bigl\{\bar U_n'(J^{11}-J_{11}^{-1})\bar U_n
	 +2\bar U_n'J^{12}\bar V_n+\bar V_n^2J^{22}\bigr\}-2 \cr
	&=n(\bar V_n-J_{21}J_{11}^{-1}\bar U_n)^2\kappa^2-2 \cr	
	&\rightarrow_d[\normal\{-\delta/\kappa^2,1/\kappa^2\}]^2\kappa^2-2
		=\chi^2_1(\delta^2/\kappa^2)-2. \cr} \eqno(4.2)$$
The probability that AIC prefers the narrow model over the wide model 
is therefore approximately $\Pr\{\chi^2_1(\delta^2/\kappa^2)\le2\}$. 
In particular, if the narrow model is perfect, the probability is 0.843, 
% pr(wide) = 0.157,
and in the border-line case suggested by this article,
i.e.~$\delta=\kappa$, the probability is 0.653.
% pr(wide) = 0.347.
See 5I below for calculations where the wide model has $q$ parameters
more than the narrow model. 

It is also instructive to see that 
${\rm AIC}_{\rm wide}-{\rm AIC}_{\rm narr}$ above is asymptotically 
equivalent to $Z_n^2-2$, where 
$Z_n=\sqrt{n}(\hatt\gamma-\gamma_0)/\hatt\kappa\rightarrow_d
\normal\{\delta/\kappa,1\}$. This is the test statistic also 
discussed in 4B. The Akaike criterion is also related to a certain 
pre-test strategy discussed in Section 5. 
The implicit advice of Section 3 would be to use the wide model
when $Z_n^2>1$ whereas the Akaike method has $Z_n^2>2$ as criterion.
It is important to note that all of these pre-test strategies
are `inadmissible' in the decision-theoretic sense, however; 
each can be uniformly improved upon, see 6(iii) below.

Akaike's criterion has a reputation for overfitting too often,
and researchers often use a more stingy criterion due to Schwarz
and others. It penalises the maximised twice log-likelihood with the factor 
$\log n$ times the number of parameters in the model, 
i.e.~subtracts $(\log n)p$ and $(\log n)(p+1)$ instead. 
The reasoning above, applied to this alternative criterion, 
shows that the Schwarz method chooses the narrow model, 
with probability tending (but slowly) to 1. 
An approximation is mentioned in 5I. 
The alternative model must be at least $\delta(\log n)^{1/2}/\sqrt{n}$
away to interest Schwarz {\csc [sic]}. 

\subsection
{\sl 4E. Evaluation of $\kappa$ through stochastic simulation.} 
The examples of Section 7 show that it is possible to compute
$J_{\rm wide}$ and $\kappa^2$ explicitly 
even for somewhat complicated departure models, in effect because
the computations only need to be carried out at the null model.
In some situations it might be too difficult, however.
One way out is then to write down the difficult elements of the 
$J_{\rm wide}$ matrix in terms of integrals, involving the
null density $f(y,\theta_0,\gamma_0)$ as well as $U(y)$ and $V(y)$,
and then carry out numerical integration. This is feasible since
only one-dimensional integrals are involved. 
This method gives a numerical value of $\kappa$ for specified 
basis point $\theta_0$. 

Another way is through stochastic simulation. Several options 
can be considered. 
(i) Simulate a large number of 
$Y_i$'s from the null distribution at some target point $\theta_0$, 
and compute score functions $U(Y_i)$ and $V(Y_i)$ along the way (see (2.3)).
Then compute empirical covariances and variances to get $J_{\rm wide}$. 
(ii) Keep $n$ fixed, simulate $Y_1^*,\ldots,Y_n^*$ from 
the null density, at some desired $\theta_0$, and compute
the estimates $\hatt\theta^*$ and $\hatt\gamma^*$ 
based on this pseudo-sample. Do this a large number of times, 
and the empirical covariance matrix for $(\hatt\theta^*,\hatt\gamma^*)$
is $J_{\rm wide}^{-1}/n$. (iii) Or drop $\hatt\theta^*$ and just evaluate
the empirical standard deviation of 
$\sqrt{n}(\hatt\gamma^*-\gamma_0)$, which is $\kappa$. 
This is a feasible approach in complex regression models,
or in parametric and semiparametric 
survival data models with censoring, 
where analytical expressions for $\kappa^2$ cannot be found. 

\subsection
{\sl 4F. Good models and dangerous departures.}
Which departures from a given narrow model are 
dangerous, and which are insignificant? 
And what qualities should a `good and robust' model have? 

We have demonstrated that the narrow model can tolerate 
$\delta=\sqrt{n}(\gamma-\gamma_0)$ up to the limit $\kappa$ in
absolute value. The numerical value of $\kappa$ depends on the
scale used, however. The appropriate scale invariant 
tolerance measure is $d=\kappa^2J_{22}=J^{22}J_{22}$, 
as is also suggested by the distances considered in 4C. 
Two numbers of this kind can be directly compared for two 
specifically envisaged model departures. A model departure
with large $d$ is less dangerous 
than one with small $d$. 

A model deviance can be studied in terms of 
$V(y)=\dell\log f(y,\theta_0,\gamma_0)/\dell\gamma$,
see (2.3). How well is $V(y)$ explained by the existing
model, represented by $U(y)$? A natural measure is the 
so-called maximal correlation,
$\rho^2\{U,V\}$, the maximal value of 
${\rm corr}\{a_1U_1+\cdots+a_pU_p,V\}^2$
as $a=(a_1,\ldots,a_p)'$ varies. 
It is well known and just a piece of linear algebra to prove that
$a_0=J_{11}^{-1}J_{12}$ maximises, with resulting
$$\rho^2\{U,V\}=J_{12}'J_{11}^{-1}J_{12}
	=1-1/(\kappa^2J_{22})=1-1/d. \eqno(4.3)$$
This invites a geometrical interpretation for the tolerance limit $d$.
The smallest possible value for $d$ is 1, which happens when
the model departure is `completely new' and orthogonal to the
existing model, with $J_{12}=0$. Only a mild departure 
in this direction can be tolerated.
So a dangerous departure is one that can be realistically 
suspected, in the first place, 
and which has a small $d$, or a small correlation.
A non-critical departure is one that has a large tolerance $d$,
or a large correlation, or one that 
perhaps is unrealistic a priori. --- A good and robust model, 
therefore, is one where
realistically suspected deviances have large tolerances $d$. 
See the examples of Section 7. 

\subsection
{\sl 4G. Can we de-bias?} 
We have demonstrated that narrow estimation,
which means introducing a deliberate bias to reduce variability,
leads to better estimator precision in a certain radius around
the narrow model. The precise quantitative result is that 
$\sqrt{n}(\hatt\mu_{\rm narr}-\mu_{\rm true})$ tends to
$\normal\{b\delta,\tau_0^2\}$, see Section 3. 
Can we remove the bias and do even better?

About the best we can do in this direction is to use
$\hatt\mu_{\rm db}=\hatt\mu_{\rm narr}-b(\hatt\gamma-\gamma_0)$.
Analysis reveals, working from the basis result 
(5.2) of the next section, 
that $\sqrt{n}(\hatt\mu_{\rm db}-\mu_{\rm true})$ 
tends to $\normal\{0,\tau_0^2+b^2\kappa^2\}$. So the bias can
be removed, but the price one pays amounts exactly to 
what was won by deliberate biasing in the first place, 
and the de-biased estimator is equivalent to $\hatt\mu_{\rm wide}$. 
The reason for the extra variability is that 
no consistent estimator exists for $\delta$. 

\subsection
{\sl 4H. Dwindling confidence.} 
%	--- Nils: check use of `svinnende' here --- 
We have established that $\hatt\mu_{\rm narr}$ 
%	narrow estimators
has higher precision than $\hatt\mu_{\rm wide}$
%	wide estimators,
for moderate misspecifications of the narrow model.
But what with further inference? 

Consider confidence intervals. The usual approximate 90\% 
interval for $\mu$ based on narrow model assumptions is 
${\rm CI}_{\rm narr}=\hatt\mu_{\rm narr}\pm1.645\,\hatt\tau_0/\sqrt{n}$,
where $\hatt\tau_0$ is consistent for $\tau_0$ of (3.2). 
But in the present local misspecification framework
$\sqrt{n}(\hatt\mu_{\rm narr}-\mu_{\rm true})$ tends to
$\normal\{b\delta,\tau_0^2\}$, and the bias destroys the 
90\% property. The probability that $\mu_{\rm true}$ is covered 
by ${\rm CI}_{\rm narr}$ converges to 
$\Pr[-1.645\le \normal\{b\delta/\tau_0,1\}\le 1.645]$.
This is always {\it strictly less than 90\%}, 
unless the narrow model is exactly true or $b$ of (3.2) is zero. 
Yes, I am shocked. 
%	Really, Nils?  
The difference is not necessarily dramatic, 
in that the coverage probability is 
above 85\% when $|b\delta|/\tau_0$ is smaller than 0.54 
and above 80\% when the ratio is smaller than 0.77.
What is important is that the narrow model based interval
always underestimates the confidence, 
under any model departure from any given parametric model, 
and that we have an illuminating explicit
formula for the true (asymptotic) coverage probability. 
%	Nils: still stronger words? 

It is not possible to remove the bias and still get a shorter
honest 90\% interval than 
${\rm CI}_{\rm wide}=\hatt\mu_{\rm wide}
	\pm1.645\,\hatt\tau_{\rm wide}/\sqrt{n}$.
This follows from analysis similar to that in 4G. 
Thus, in a way, within the chosen large-sample framework,
and provided we insist on guaranteed levels,  
we cannot carry out confidence and testing analysis 
better than with wide model methods, despite the fact
that narrow estimators often have better precision than wide ones.
A practical proposal is to use $\hatt\mu_{\rm narr}$
%	the narrow model based estimator,
when theory and analysis suggest that it is more precise than
$\hatt\mu_{\rm wide}$,
%	the the wide model based estimator, 
but to supplement it with
a confidence interval obtained through nonparametric or
wide-model-parametric bootstrapping. 
The point is to obtain an honest 90\% interval, for example,
built around $\hatt\mu_{\rm narr}$. 

Let us finally point out that narrow based intervals 
in some natural ways perform better than wide model ones, 
under mild misspecifications,
since they are, indeed, narrower. 
Assume the loss incurred by using CI to cover $\mu$ is of the form
$$L[(\theta,\gamma),{\rm CI}]
	=I\{\mu(\theta,\gamma)\notin{\rm CI}\}
		+\sqrt{n}w\,{\rm length(CI)},$$
where $w$ is an appropriately chosen weight factor.
The idea is to combine the two desiderata of confidence intervals
into one measure; they should miss with low probability and have
short length. The asymptotic risk functions for
${\rm CI}_{\rm narr}=\hatt\mu_{\rm narr}\pm z_0\hatt\tau_0/\sqrt{n}$
and ${\rm CI}_{\rm wide}=\hatt\mu_{\rm wide}\pm z_1\hatt\tau/\sqrt{n}$,
under model $P_n$, become
$$\eqalign{
{\rm risk}_{\rm narr}&=\Pr\bigl[|\normal\{b\delta/\tau_0,1\}|\ge z_0\bigr]
   	+2wz_0\tau_0, \cr
{\rm risk}_{\rm wide}&=\Pr\bigl[|\normal\{0,1\}|\ge z_1\bigr]
	+2wz_1(\tau_0^2+b^2\kappa^2)^{1/2}. \cr} \eqno(4.4)$$
Again the best narrow method will be better than the best wide method,
for moderate deviances $\delta$ from zero.

\subsection
{\sl 4I. Deviances in several directions.} 
Our results can be generalised to a framework with two or more 
types of departure from the basic model, like both quadraticity and 
variance heterogeneity in regression. See 5I. 

\bigskip
{\bf 5. Classes of compromise estimators.}
%	 and a general performance study} 
We have so far concentrated on $\hatt\mu_{\rm narr}$ and 
$\hatt\mu_{\rm wide}$ to estimate 
$\mu=\mu_{\rm true}=\mu(\theta_0,\gamma_0+\delta/\sqrt{n})$.
These rather cyclopic estimators can however be combined to form 
dimeric ones that perhaps work well both under
the null model and the local alternative. 
This section considers and develops 
various more complex estimators with this aim. 
Some key words indicating the different types 
that will be discussed are 
pre-test or if-else estimators, 
mixture or weighted estimators, 
Bayes and empirical Bayes estimators,
minimax estimators,
the Bayesian epsilon estimator,
and limited translation estimators.

Comparing all of these approaches may appear 
to be a formidable task, since the problem conceivably depends  
upon the particularities of the narrow model, 
the type and degree of deviance from it, 
and the specific parameter estimand under study.
The comparison problem can however be drastically reduced,
as we show in 5D below. Each of a large class of estimators 
for $\mu_{\rm true}$ has a cousin which estimates
$a$ in a $\normal\{a,1\}$ situation with one observation 
under squared error loss! The underlying 
one-one correspondence makes it possible to study
the performance of general estimation approaches
rather simply and rather generally,
and this is indeed done in Section 6. 

\subsection
{\sl 5A. If-else of pre-test estimators.}
`The responsibility of tolerance lies with those who have 
the wider vision.'  
% 	Nils: Should I use quotation marks? 
A procedure that is sometimes advocated in model choice problems
and which perhaps is consistent with George Eliot's view 
is as follows, in the present context: Test the hypothesis
$\gamma=\gamma_0$ against the alternative $\gamma\not=\gamma_0$, say at
the 10\% level; if accepted, then use $\hatt\mu_{\rm narr}$,
if rejected, then use $\hatt\mu_{\rm wide}$. 
Choosing the $Z_n^2=n(\hatt\gamma-\gamma_0)^2/\hatt\kappa^2$ test
also discussed in 4B, this suggestion amounts to 
$$\hatt\mu_{\rm pre}=\hatt\mu_{\rm narr}I\{Z_n^2\le 1.645^2\}
	+\hatt\mu_{\rm wide}I\{Z_n^2> 1.645^2\},
	\quad 1.645^2={\rm upper\ 10\%\ point\ of \ }\chi^2_1. \eqno(5.1)$$
But this method sticks too rigidly to the narrow model. 
The theory of Section 3 suggests that one should rather use 
the much smaller value 1 as cut-off point, since 
$|\delta|\le\kappa$ corresponds to $n(\gamma-\gamma_0)^2/\kappa^2\le 1$,
and $Z_n^2$ estimates this ratio. 
Using 1 as cut-off corresponds to a much more relaxed significance level, 
indeed to 31.7\%, which in this sense becomes the optimally chosen 
significance level. 
The Akaike method corresponds to using 2 as cut-off point for $Z_n^2$
with significance level 15.7\%, see 4D.  
%	Nils: satisfied? In what sense is this d-value optimal?
%	Note my pretest_risk_maximum.data file.   
Observe that $\sqrt{n}(\hatt\mu_{\rm pre}-\mu_{\rm true})$ 
tends to a mixture of two normals, as further commented upon below. 

\subsection
{\sl 5B. Mixture estimators.}
Another natural idea is 
$\hatt\mu_{\rm lin}=(1-c)\hatt\mu_{\rm narr}+c\hatt\mu_{\rm wide}$.
To find the approximate distribution of this estimator it is 
necessary to go somewhat beyond the basic proposition of Section 2,
in that the simultaneous limit distribution of the narrow and the wide
estimators is needed. This can be found by studying the proof of
the proposition, however. 
Utilising (2.4) and (2.5) it follows via some analysis that
$$\eqalign{
\sqrt{n}(\hatt\mu_{\rm narr}-\mu_{\rm true})
	&\rightarrow_d b\delta+(\delltheta)'J_{11}^{-1}M, \cr
\sqrt{n}(\hatt\mu_{\rm wide}-\mu_{\rm true}) 
	&\rightarrow_d \mtrix{\delltheta \cr \dellgamma \cr}'
		J_{\rm wide}^{-1}\mtrix{M \cr N\cr}, \cr
Z_n=\sqrt{n}(\hatt\gamma-\gamma_0)/\hatt\kappa
	&\rightarrow_d Z=(\delta+J^{21}M+J^{22}N)/\kappa, \cr} \eqno(5.2)$$
in which $(M,N)\sim \normal_{p+1}\{0,J_{\rm wide}\}$. 
The convergence is simultaneous, and takes place under the 
$P_n$ sequence of models (2.2). 
Note that $Z\sim \normal\{\delta/\kappa,1\}$. 
%% \eject 

Now the limit distribution of $\hatt\mu_{\rm lin}$ 
can be obtained. The result is 
$$\sqrt{n}(\hatt\mu_{\rm lin}-\mu_{\rm true})
	\rightarrow_d (1-c)b\delta+(1-c)(\delltheta)'J_{11}^{-1}M
	+c\mtrix{\delltheta \cr \dellgamma \cr}'
		J_{\rm wide}^{-1}\mtrix{M\cr N \cr}.$$
This is a normal distribution with mean value $(1-c)b\delta$,
and with some stamina its variance is found to be $\tau_0^2+c^2b^2\kappa^2$, 
in the notation of Section 3. 
The ideal value of $c$ that minimises the asymptotic mean squared error
%	Nils: clear up this very slight thing
for $\hatt\mu_{\rm lin}$ is 
$c_0=\delta^2/(\kappa^2+\delta^2)=a^2/(1+a^2)$,
featuring the key quantity $a=\delta/\kappa$.
The accompanying minimum value
%	$n\,E(\hatt\mu_{\rm mix}-\mu_{\rm true})^2$ 
is equal to $b^2\kappa^2 a^2/(1+a^2)+\tau_0^2$. 
Note that this is always better than both the $b^2\kappa^2+\tau_0^2$ 
achieved by $\hatt\mu_{\rm wide}$ and the $b^2\delta^2+\tau_0^2$
achieved by $\hatt\mu_{\rm narr}$.  	      

The problem is of course that $c_0$ is unknown since $\delta$ is.
Using the empirical counterpart of $\delta=\sqrt{n}(\gamma-\gamma_0)$ 
invites $Z_n=\sqrt{n}(\hatt\gamma-\gamma_0)/\hatt\kappa$ 
to be inserted for $\delta/\kappa$, i.e.~$Z_n^2$ estimates $a^2$,
and one could try out the diophthalm
%	Nils: --- something with two eyes --- 
$$\hatt\mu_{\rm eb}={1\over 1+Z_n^2}\hatt\mu_{\rm narr}
	+{Z_n^2\over 1+Z_n^2}\hatt\mu_{\rm wide}. \eqno(5.3)$$
Note the Steinean overtones. 
The empirical Bayes connection that gives its subscript 
is noted in 5F below. 

\subsection
{\sl 5C. Compromise estimators.}
%	{\sl 5C. Bilingual estimators.}
Let us generalise. 
We shall be content to study estimators in the 
fairly large class of {\it compromise estimators},
which are bilingual and want the best from two worlds, 
and which we describe as follows. Its prime members are of the type
$$\mu^*=\{1-c(Z_n)\}\hatt\mu_{\rm narr}
		+c(Z_n)\hatt\mu_{\rm wide},\eqno(5.4)$$
where $c(z)$ is almost everywhere continuous. Note that 
the previously considered estimators are of this form. 
The additional members that are included are those that 
can be closely approximated by (5.4) type ones by linearisation.
More specifically, the limit distribution result (5.5) below is
required to hold. It suffices for $\mu^*$ to be of the form
$m(\hatt\mu_{\rm narr},\hatt\mu_{\rm wide},Z_n)$ for 
some smooth function $m(\mu_1,\mu_2,z)$ with the property
that $m(\mu,\mu,z)\equiv\mu$. 
%	Nils: In such cases ... takes the role of c(z) and ...
An example is the harmonic 
variety $\exp\bigl[\{1-h(Z_n)\}\log\hatt\mu_{\rm narr}
	 +h(Z_n)\log\hatt\mu_{\rm wide}\bigr]$ (which
can be used in cases where $\mu$ is positive).

\subsection
{\sl 5D. Comparison of estimators: a drastic reduction.} 
We wish to study the performance of all these estimators,
and to compare pairs of them, 
w.r.t.~the limiting mean squared error criterion.

\smallskip
{\csc Theorem.} {{\sl
The compromise estimator (5.4) has limit distribution, 
under $P_n$ of (2.2), given by 
$$\sqrt{n}(\mu^*-\mu_{\rm true})
	\rightarrow_d
	\Lambda=\{1-c(Z)\}\{b\delta+(\delltheta)'J_{11}^{-1}M\}
	+c(Z)\mtrix{\delltheta \cr \dellgamma \cr}'
		J_{\rm wide}^{-1}\mtrix{M \cr N \cr}. \eqno(5.5)$$
The mean squared error of the limit distribution can be written as
$$\E\Lambda^2=b^2\kappa^2\,E\bigl\{\delta/\kappa-c(Z)Z\bigr\}^2
	+\tau_0^2=b^2\kappa^2R(\delta/\kappa)+\tau_0^2, \eqno(5.6)$$
in which 
$$R(a)=E\bigl\{c(Z)Z-a\bigr\}^2
	{\sl \ and\ }Z\sim \normal\{a,1\}. \eqno(5.7)$$
\smallskip}}

{\csc Proof:} 
(5.5) follows from (5.2) and the continuous mapping theorem
of weak convergence. 
To characterise this limit variable $\Lambda$,
study its distribution conditional on $Z=z$.
Ordinary techniques from multivariate analysis,
working from (5.2), lead to 
$$\mtrix{M \cr N \cr}\midd \{Z=z\}\sim
	\normal_{p+1}\{\mtrix{0 \cr (\kappa z-\delta)/\kappa^2 \cr},
 	\mtrix{J_{11} &J_{12} \cr
		J_{21}  &J_{22}-1/\kappa^2 \cr}\}.$$
Several algebraic and multivariate details later one arrives at 
$$\Lambda\midd \{Z=z\}\sim \normal\{b\delta-c(z)b\kappa z,\tau_0^2\},
	\quad {\rm where\ } Z\sim \normal\{\delta/\kappa,1\}. $$
Expression (5.6) for the limiting mean squared error can now 
be worked out, studying first the $z$-conditional and then 
the unconditional mean value of $\Lambda^2$. \square

\smallskip
This result contains those associated with (3.1) and (3.2) 
as well as (5.1) and the case of fixed $c$ studied above.  
Observe that the unconditional distribution of $\Lambda$ 
is non-normal unless $c(z)$ is constant in $z$. 
Note also that the unfamiliar type of limit distribution 
is not a peculiarity of the chosen local neighbourhood asymptotics, 
since $\Lambda$ is typically non-normal even in the null model case. 

A particular consequence of the theorem is that 
{\it it suffices to compare different versions 
of the function} $R(a)$,
as a function of $a=\delta/\kappa$, since $b\kappa$ and $\tau_0$ 
remain unchanged for different estimators.
%	well? Nils? 
(We disregard the rather simple cases in which $b=0$, 
see `case (i)' of Section 3's Result, 
under which all compromise estimators
have $\normal\{0,\tau_0^2\}$ as limit distribution.)
This constitutes an impressive reduction of the 
original comparison problem. 
Note that $R(a)$ is simply the risk function for 
the estimator $c(Z)Z$ for $a$ 
in the one-observation $Z\sim \normal\{a,1\}$ problem under squared error loss. 
There is a simple one-to-one correspondence 
from general compromise estimators to estimators 
$\hatt a(Z)$ of $a$ based on $Z$, via
$$\hatt a(z)=c(z)z, \quad c(z)=\hatt a(z)/z. \eqno(5.8)$$	
We stress the generality: A comparison between
the four natural estimators $\hatt\mu_{\rm narr}$,
$\hatt\mu_{\rm wide}$, $\hatt\mu_{\rm pre}$ of (5.1), 
and $\hatt\mu_{\rm eb}$ of (5.3),
for example, can be carried out entirely in the realm
of the classical $Z\sim \normal\{a,1\}$ situation,
by simply drawing the four $R(a)$ curves. 
See Section 6 for examples. 
And the conclusions from this comparison 
remain correct and relevant in {\it every} 
`moderate misspecification' problem,
cf.~the wide span of problems that Examples A--G represent.
Finally one is allowed to go the other way: 
{\it Your favourite estimator for $a$} in the $\normal\{a,1\}$ problem 
(where $a$ may be rumoured to be in the vicinity of zero) 
can be transported to a useful estimator for 
any given estimand in any given moderate misspecification situation.  

In most cases it holds that 
$c(z)=c(-z)$, implying $R(a)=R(-a)$,
making it necessary to study only non-negative $a$'s. 
The parameter $a$ measures the degree of misspecification 
from the narrow model. The important range is perhaps
$[-4,4]$, where $a=0$ means correctness of the narrow model,
$a=\pm1$ are the turning points after which the wide estimator
becomes better than the narrow one, and values beyond $\pm3$
could be thought of as clearly detectable departures from the
narrow model, cf.~power considerations (4.1), (4.2). 
(The 5\% level test has power 0.851 and 0.979 at $a=3$ and $a=4$,
whereas the 10\% level test has 0.912 and 0.991 at the same points.)
%	--- Nils:
%	delta		power		power
%	0		0.050		0.100
%	kappa		0.170		0.264
%	2kappa		0.516		0.639
%	3kappa		0.851		0.912
%	4kappa		0.979		0.991

These remarks also illustrate the importance of 
thinking about prior information related to the parameter $a$,
for example its possible range. In Example B,
studied in Sections 1 and 7 and in Hjort (1993), 
$a$ must be non-negative a priori,
and in other cases it could be natural to restrict attention
to the $[-4,4]$ range, say, or to postulate a prior
density for $a$. Such a prior could reflect serious prior
beliefs, in the Bayesian fashion, or be used as a 
mathematical device to reach an estimator with minimum
possible averaged mean squared error. Objectivists  
fretting at such ideas should note that the two  
classical solutions here, $\hatt\mu_{\rm narr}$
and $\hatt\mu_{\rm wide}$, correspond to full faith
in the priors $I_0$ and $1$, respectively, where
$I_0$ is the degenerate distribution at zero and 
$1$ is the flat non-informative prior for $a$.   
This is made clear in the course of 
the two following subsections, 
where the correspondence between moderate misspecification 
problems and the $\normal\{a,1\}$ situation is explained
also for Bayesian matters.  

\subsection
{\sl 5E. Prior and posterior distributions for $a$.} 
One is used to seeing that `the prior is washed out by the data'.
Assume for example that 
a prior density $p_0(\theta,\gamma)$ is placed on $(\theta,\gamma)$,
with resulting Bayes estimators 
$(\hatt\theta_B,\hatt\gamma_B)$, expected values in the posterior
density $p_0(\theta,\gamma\midd {\rm data})$. Then these are 
typically asymptotically equivalent to the ML estimators, 
in the precise sense that 
$\sqrt{n}(\hatt\theta-\hatt\theta_B)\rightarrow_p0$ 
and $\sqrt{n}(\hatt\gamma-\hatt\gamma_B)\rightarrow_p0$,
in the frequentist framework $P_n$.  
%	corresponding to $f(x)=f(x,\theta_0,\gamma_0
%	+\delta/\sqrt{n})$ being the true model
%	for data $X_1,\ldots,X_n$. 
This is a fairly standard result under
null model conditions, and the more delicate case of $\delta\not=0$
can be treated using methods in Hjort (1986b). 

This result uses a fixed prior for $(\theta,\gamma)$, and 
is somewhat irrelevant in the present context of moderate
misspecification. It appears more natural to operate with 
a fixed prior for 
$(\theta,\delta)=(\theta,\sqrt{n}(\gamma-\gamma_0))$,
or, equivalently, a fixed prior $p(\theta,a)$ for
$(\theta,a)=(\theta,\sqrt{n}(\gamma-\gamma_0)/\kappa)$. 
We think of the prior distribution for $a$ as reflecting
prior beliefs about the suitability of the narrow $f(y,\theta,\gamma_0)$ 
model, cf.~the discussion above. 

In this situation the prior information regarding $\theta$ 
will still be overwhelmed by the data, but not the part related to $a$.
Information about $a$ lies in 
$Z_n=\sqrt{n}(\hatt\gamma-\gamma_0)/\hatt\kappa$, 
which is not consistent,
but has a limiting variable $Z\sim \normal\{a,1\}$. 
Intuitively, therefore, the posterior density $p(a\midd {\rm data})$
should for large $n$ simply be close to 
$p(a\midd z)$ in the situation where $Z$ is $\normal\{a,1\}$ and   
$a$ has prior proportional to $p(\theta_0,a)$.
To prove it, let us study $p(a\midd Y_1,\ldots,Y_n)$ 
when $n$ grows, under the $P_n$ model, where 
$f(y)=f(y,\theta_0,\gamma_0+\delta_0/\sqrt{n})$ for
some fixed values of $\theta_0$, $\delta_0$.  
Let $L_n(\theta,\gamma)=\prod_{i=1}^nf(Y_i,\theta,\gamma)$ 
be the $n$'th likelihood. By judicious second order Taylor
expansion analysis it can be established that 
$$H_n(s,t)={L_n(\hatt\theta+s/\sqrt{n},\hatt\gamma+t/\sqrt{n}) 
	    \over
	    L_n(\hatt\theta,\hatt\gamma)}
	    \rightarrow_d
	    H(s,t)=\exp\Bigl\{-{1\over2}\mtrix{s \cr t \cr}'
			J_{\rm wide}\mtrix{s \cr t \cr}\Bigr\}$$
under $P_n$ of (2.2). 
The convergence takes place in each Skorokhod space
%	Nils: spell it in Russian 
%	{\cyr Skorokhod} ? 
$D[-A,A]^{p+1}$. Let now $g(\theta,a)$ be any bounded function.
Then one may deduce
$$\eqalign{
\E\{g(\theta,a)\midd {\rm data}\}
 &={\int\int g(\theta,\sqrt{n}(\gamma-\gamma_0)/\kappa)
	L_n(\theta,\gamma)
	p(\theta,\sqrt{n}(\gamma-\gamma_0)/\kappa)
	\sqrt{n}/\kappa\,\d\theta\,\d\gamma
	\over
	\int\int L_n(\theta,\gamma)
	p(\theta,\sqrt{n}(\gamma-\gamma_0)/\kappa)
	\sqrt{n}/\kappa\,\d\theta\,\d\gamma} \cr
&={\int\int g(\hatt\theta+s/\sqrt{n},Z_n'+t/\kappa)
	H_n(s,t)p(\hatt\theta+s/\sqrt{n},Z_n'+t/\kappa)\,\d s\,\d t
	\over
	\int\int H_n(s,t)p(\hatt\theta+s/\sqrt{n},
		Z_n'+t/\kappa)\,\d s\,\d t} \cr
&\rightarrow_d
{\int\int g(\theta_0,Z+t/\kappa)H(s,t)p(\theta_0,Z+t/\kappa)\,\d s\,\d t
	\over
	\int\int H(s,t)p(\theta_0,Z+t/\kappa)\,\d s\,\d t} \cr
&={\int g(\theta_0,Z+t/\kappa)\exp(-\half t^2/\kappa^2)\pi(Z+t/\kappa)\,\d t 
	\over
	\int \exp(-\half t^2/\kappa^2)\pi(Z+t/\kappa)\,\d t} \cr
&={\int g(\theta_0,a)\exp\{-\half(Z-a)^2\}\pi(a)\,\d a 
	 \over
	 \int \exp\{-\half(Z-a)^2\}\pi(a)\,\d a}\,,\cr} $$
in which $\pi(a)={\rm const.}\,p(\theta_0,\delta)$ is the prior 
for $a$ given the information $\theta=\theta_0$,
and $Z_n'=(\hatt\kappa/\kappa)Z_n$ was used for notational simplicity.
The necessary mathematical details have to do with 
(i) securing convergence inside $[-A,A]^{p+1}$,
utilising the proposition of Section 2, along with (5.2);
(ii) 
% cleverly 
carrying out a certain inner $p$-dimensional normal integration;   
and (iii) bounding integrands outside $[-A,A]^{p+1}$ for large $A$.
The arguments that are needed  
resemble those explained in Hjort (1986b)
(to reach a different conclusion,
in a different problem), and are left out here. 

By considering $g=g(a)$ above it is clear that 
$$\pi_n(a\midd {\rm data})\rightarrow_d
	\pi(a\midd Z)={\phi(Z-a)\pi(a) \over 
	\int\phi(Z-a)\pi(a)\,\d a}, \eqno(5.9)$$ 
under $P_n$, where $Z\sim \normal\{a,1\}$ is as in (5.2).   
This is what was predicted above. 
If $a$ has some prior distribution $\d\pi(a)$ that perhaps does not
have a density, then the arguments can be repeated to 
give $\d\pi_n(a\midd {\rm data})\rightarrow_d
	{\rm const.}\,\phi(Z-a)\,\d\pi(a)$ instead.

\subsection
{\sl 5F. Bayes and empirical Bayes estimators.} 
We should distinguish between kosher Bayes and approximate
Bayes estimators. A prior density $p(\theta,a)$ 
for $(\theta,a)$ leads to the exact Bayes solution 
$\hatt a_n=\E\{a\midd Y_1,\ldots,Y_n\}$. This is usually a very complicated
expression, and in view of (5.9) it is tempting to 
work directly in the limit distribution and use $\hatt a(Z_n)$
instead, where 
$$\hatt a(z)=\E\{a\midd Z=z\}=
	{\int a\phi(z-a)\pi(a)\,\d a
	 \over
	 \int \phi(z-a)\pi(a)\,\d a}
	=z+{\dell\over \dell z}\log\int\phi(z-a)\pi(a)\,\d a. \eqno(5.10)$$
But the arguments of 5E can be used to reach
$$\hatt a_n=Z_n'
	+{\dell\over \dell z}\log\int
		\phi(Z_n'-a)\pi(a)\,\d a+O_p(1/\sqrt{n}),$$
where again $Z_n'=(\hatt\kappa/\kappa)Z_n$. This proves
$\hatt a_n-\hatt a(Z_n)\rightarrow_p0$, under $P_n$, allowing
us to use $\hatt a(Z_n)$ instead when we devise and study
Bayes solutions in our large-sample framework. In particular 
we do not have to bother with the part of the prior information
that has to do with $\theta$. 

Some specific Bayesian and empirical Bayesian constructions follow. 

(i) 
Suppose $a$ is $\normal\{0,\sigma^2\}$
(where the size of the spread parameter $\sigma$ matters more than
the normality). Then $Z\sim \normal\{0,\sigma^2+1\}$,
and $\hatt a(z)=\{\sigma^2/(\sigma^2+1)\}\,z$, with Bayes risk
$\sigma^2/(\sigma^2+1)$. If $Ea^2=\sigma^2$ is unknown, 
a simple guess is $Z_n^2$, since $Z_n$ estimates $a$. 
This brings forward the empirical Bayes estimate
$\hatt a_{\rm eb}(Z_n)=\{Z_n^2/(Z_n^2+1)\}\,Z_n$ for $a$. 
But this corresponds to $\hatt\mu_{\rm eb}$ of (5.3), 
explaining its empirical Bayes interpretation. 

One may also consider other estimators for $q=\sigma^2/(\sigma^2+1)$
here. Each such $\hatt q=\hatt q(Z)$ 
leads to an $a^*=\hatt a(Z,\hatt q)$,
and in its turn to a new estimator $\mu^*$ for $\mu_{\rm true}$
via (5.4) and (5.8). The fact that $EZ^2=\sigma^2+1$ suggests  
$\tilda q=(Z^2-1)_+/Z^2$, which in fact is the ML solution, 
or similar versions. 
Another proposal is to put a vague hyper prior on
$\sigma$, or directly on the ratio $q$. 
The Bayes solution becomes
$\E\{a\midd Z=z\}=\E_z\E_z\{a\midd \sigma\}=\E_z(qz)=\hatt q(z)z$,
in which $\hatt q(z)=\int_0^1 qp(q\midd z)\,dq$ is the posterior 
density of $q$ for given $Z=z$. The usual 
choice for a non-informative prior for a scale parameter
like $\sigma$ is to have $\log\sigma$ uniform. This leads
to $p(q)={\rm const.}\,\{q(1-q)\}^{-1}$ on $[\eps,1-\eps]$, say, 
for $q$, with a corresponding explicit $\hatt q(z)$.  
In fact it turns out that
$$\hatt q(z)={\int_\eps^1(1-q)^{-1/2}\exp\{-\half(1-q)z^2\}\,dq
	\over
	\int_\eps^1q^{-1}(1-q)^{-1/2}\exp\{-\half(1-q)z^2\}\,dq} 
					 	\eqno(5.11)$$
is substantially better than $\tilda q=(z^2-l)_+/(z^2+1-l)$,
where $0\le l\le 1$, for a wide right interval $(q_0,1)$ 
of $q$ values. 
Heroic numerical integrations have demonstrated this, 
via computations and comparisons of $\E_q\midd q^*-q|$ 
for the various estimators. The $\hatt q$ above, with $\eps=0.05$, 
is for example much better than the $\tilda q$ ones,
for $q$ in $(0.20,1)$. 

(ii) 
Suppose $a$ comes from $\pi_0(a)$ with probability $p_0$
and from $\pi_1(a)$ with probability $p_1$. 
Then calculations show that $\hatt a(z)$ is of the mixture form 
$w_0(z)\hatt a_0(z)+w_1(z)\hatt a_1(z)$, where
$\hatt a_j(z)$ is the Bayes estimator under theory $p_j(a)$,
and $w_j(z)=p_jh_j(z)/\{p_0h_0(z)+p_1h_1(z)\}$,
and $h_j(z)=\int\phi(z-a)\pi_j(a)\,\d a$. An interesting special case
is the prior distribution $a\sim(1-\eps)I_0+\eps \normal\{0,\sigma^2\}$,
where $I_0$ denotes the degenerate distribution at zero. This 
is a `Bayesian epsilon' approach, where the statistician 
is rather convinced of the narrow model's correctness but
allows the data to express a different opinion with probability $\eps$.
In this case
$$\hatt a(z)={\eps \over \eps+(1-\eps)B(z)}{\sigma^2\over \sigma^2+1}z,
	\quad
	B(z)={h_0(z)\over h_1(z)}
	    =\sqrt{\sigma^2+1}\exp\Bigl\{-{1\over2}
		{\sigma^2\over \sigma^2+1}z^2\Bigr\}. \eqno(5.12)$$
Again $\sigma^2$ has to be specified or estimated. 
One possibility is $\hatt\sigma^2=Z^2/\eps$, 
since $Ea^2=\eps\sigma^2$ and $Z$ estimates $a$; 
other versions can be constructed as in (i) above. 
% Using a fixed $\sigma$ gives unbounded risk, as $a$ grows,
% whereas the empirical Bayesian versions typically have bounded risk
% functions. 
% Not necessarily true, Nils. Investigate? 
%	Nils: this remark later on instead? 

(iii) If it is assumed that $|a|\le m$ a priori then 
the Bayes solution (5.10) with a uniform prior on $[-m,m]$ 
should give an estimator with good risk properties on this interval. 

\subsection
{\sl 5G. Minimax type estimators.} 
The remarks about the $a$ parameter in 5D suggest that
its range could usefully be taken to be bounded, a priori, 
in some situations. If $a$ is postulated to be in $[-m,m]$,
for example, then estimators $a^*$ exist that are uniformly
better than $z$, which means, by our basic correspondence
theorem, that estimators $\mu^*$ exist 
that are uniformly better than $\hatt\mu_{\rm wide}$. 
If in particular $a_m^*$ is a minimax estimator,
with maximum risk $r_m<1$ for $R(a)$ in $[-m,m]$, then
$\mu^*$ of (5.4), defined via (5.8), 
has a minimax property: It minimises the
limit distribution version of 
$$\max_{|\gamma-\gamma_0|\le m\kappa/\sqrt{n}}
	n\E_{\theta_0,\gamma}\{\hatt\mu-\mu(\theta_0,\gamma)\}^2$$
over all estimators $\hatt\mu$, and achieves  
$\max_{|\delta|\le m\kappa}\E\Lambda^2=b^2\kappa^2r_m+\tau_0^2<\tau^2$.   

How do such minimax estimators look like? 
It is known that $a_m^*(z)$ is the proper Bayes solution w.r.t.~a
prior distribution concentrated in a finite number of points,
see e.g.~Lehmann (1983, chapter 4.3). This least favourable
prior has been found for small values of $m$, at least for $m\le 1.5$,
and Bickel (1981) gives approximate results for $m$ large. 
We mention that $a^*=m\tanh(mz)$, the Bayes solution under
a symmetric two-point prior in $\pm m$, is minimax, provided 
$m\le 1.05$. This is relevant here since $[-1,1]$ is the range 
for $a$ where narrow estimation is better than wide estimation.   
%	The minimax risk for $m=1$ is about .45. 
Bickel shows that the distribution with density 
$\pi_m(a)=\cos^2(\half\pi a/m)/m$ for $|a|\le m$
is approximately least favourable, for large $m$. This suggests 
trying out 
$$\hatt a_{\rm bic}(z)={\int_{-m}^ma\phi(z-a)\cos^2(\half\pi a/m)\,\d a
	\over
	\int_{-m}^m \phi(z-a)\cos^2(\half\pi a/m)\,\d a}, \eqno(5.13)$$
It is {\it not} approximately minimax on $[-m,m]$, but 
it is uniformly better than $\hatt a_{\rm wide}=z$ 
in a certain interval around 0. A simpler possibility is to use the 
ML solution when $|a|\le m$ a priori, that is,
$$\hatt a_{\rm res}(z)=-m {\rm\ when\ }z\le-m, 
		\quad z {\rm\ on\ }[-m,m], 
		\quad m {\rm\ when\ }z\ge m.\eqno(5.14)$$
This is not quite as good as using the
proper minimax solution on $[-m,m]$, however.
%	... and $\hatt a_m(z)$ is not smooth

Finally we should include estimators of the Efron--Morris variety,
see Efron and Morris (1971) and Lehmann (1986, chapter 4.2). 
These aim at minimising Bayes risk, under normal priors, 
subject to having maximum risk less than some prescribed level.  
A particular case of these is pertinent here, namely the 
`limited translation estimator' 
$$\hatt a_{\rm em}(z)=z+m {\rm\ when\ }z\le-m, 
		\quad 0 {\rm\ on\ }[-m,m], 
		\quad z-m {\rm\ when\ }z\ge m. \eqno(5.15)$$ 
These come close to minimising maximum risk subject to doing well at 
$a=0$, see Bickel (1983, 1984) and Berger (1982). 
They are not smooth enough to be 
admissible. An alternative estimator which can be proposed is 
$$\hatt a_{\rm atan}(z)=z-m(2/\pi)\arctan z. \eqno(5.16)$$
It is motivated from Bickel's study of $\hatt a_{\rm em}$ 
and its connection to bounded influence functions in robust 
estimation of location, and is scaled so that it has the same maximum
risk $1+m^2$ as (5.15) (see below). 
%	{\bf Nils:} Efron--Morris (1971), Bickel (1983). 
%	Nils: Efron--Morris variety, 
%$$\hatt a(z)=\cases{z+d/(\sigma^2+1) &when $z\le -d$, \cr
%		\sigma^2/(\sigma^2+1)\,z &when $|z|\le d$, \cr
%		z-d/(\sigma^2+1)	&when $z>d$, \cr}$$ 

\subsection
{\sl 5H. Some concluding comments.}

(i) 
Observe the generality under which the comparisons 
of Sections 5 and 6 are made. 
They are valid and relevant for all of Examples A--G 
(with appropriate modifications for case B, see Hjort (1993)) 
and for all parameter estimands, via (5.4)--(5.7). 

(ii) 
When applied to a particular estimand in a particular model 
these comparisons should perhaps also include 
the nonparametric contender. In Example A, for example, 
one could compare the wide and the narrow parametric methods to
the sample median. 

(iii) 
We have been motivated by approximate mean squared error
$\E_{P_n}(\mu^*-\mu_{\rm true})^2$ when comparing estimator
performance. We haven't quite worked with the limit of 
$n$ times the mean squared error, but rather with $\E\Lambda^2$ 
in (5.6), using the limit distribution. 
This is both easier and more meaningful. 
This is a minor technical point, however; usually the two agree.
See Lehmann's (1983) Lemma 5.1.2, for example. 
%	see Lehmann (1983, p.~341 and 466). 

(iv)
Our study has been a large-sample one, and its relevance for
finite samples hinges on the degree to which 
$n\E_{P_n}(\mu^*-\mu_{\rm true})^2$ approximates its limit,
and perhaps even more on how well the limit results can predict
regions of relative superiority for one estimator over another.
K\aa resen (1992) has some explicit calculations that support
conclusions from the asymptotic framework; even in cases where
the mean squared error convergence is slow the points at which
two risk functions cross are well predicted by the limit calculations.

(v) 
One might wish to study $L_1$ error $\sqrt{n}\E_{P_n}|\mu^*-\mu_{\rm true}|$
and its limit distribution version $\E|\Lambda|$ instead. 
There is a parallel result to (5.6) and (5.7) for this problem.
Let $L(x)$ be the function $\E|x+\normal\{0,1\}|=x+2\phi(x)-2x\{1-\Phi(x)\}$.
Then $\mu^*$ of (5.4) has 
$$\E|\Lambda|=\tau_0\int 
	L\bigl((b\kappa/\tau_0)\,(\delta/\kappa-c(z)z)\bigr)\,
			\phi(z-a)\,\d z
	=\tau_0\E_aL\bigl(\rho(c(Z)Z-a)\bigr), \eqno(5.17)$$
letting $a=\delta/\kappa$ again and $\rho=|b|\kappa/\tau_0$. 
There is once more a correspondence between 
compromise estimators $\mu^*$ of $\mu$ 
and estimators $\hatt a(z)=c(z)z$ of $a$, but the 
$L_1$ loss function $|\mu^*-\mu|$ for $\mu$ is transformed 
to loss function $L(\rho(\hatt a-a))$ for $a$.  
And there is still a `tolerance radius' 
around the narrow model inside of which
misspecification is favourable, but one does not get the 
clear-cut $|\delta|\le\kappa$ answer. 
The narrow and the wide procedures have respectively 
$\E|b\delta+\tau_0N|$ and $(b^2\kappa^2+\tau_0^2)^{1/2}\E|N|$ 
as limiting risks, where $N\sim \normal\{0,1\}$. 
The tolerance radius becomes in fact 
$|\delta|\le a_0\kappa=a_0(\rho)\kappa$,
or $|a|\le a_0(\rho)$, where $|a|\le a_0$ corresponds to 
$\E|\rho a+N|\le(1+\rho^2)^{1/2}\,\E|N|$. 
Computations show that $a_0(\rho)$ starts 
at $1.00$ for $\rho=0$ and slouches towards $\sqrt{2/\pi}=0.7979$ 
as $\rho$ grows. The $L_1$ criterion for estimation of $\mu$ 
accordingly tolerates slightly less misspecification 
than the $L_2$ criterion.  

\subsection
{\sl 5I. Model departures in several directions.}
This article has focussed on the case of a single extraneous
parameter to describe deviation from a model. 
In many situations it is worthwhile to study two or more 
types of model departure simultaneously, like both quadraticity 
and variance heterogeneity in regression. 
Most of our results can be generalised to such a situation,
some with ease and some requiring harder work. 
A very brief outline is given here. 
 
Suppose $f(y,\theta_0,\gamma_0+\delta/\sqrt{n})$ is the true model,
where $\gamma$ and $\delta$ are $q$-dimensional. 
The natural criterion for when {\it each narrow estimator} 
is asymptotically more precise than
its wide contender becomes $\delta\delta'\le J^{22}$, where
$J^{22}=(J_{22}-J_{12}'J_{11}^{-1}J_{12})^{-1}$ is $q\times q$. 
This describes an ellipse or ellipso\"\i d around the null model,
and is in fact equivalent to $\delta'(J^{22})^{-1}\delta\le1$,
generalising (3.4). 
The border line, the crossing of which means coming into 
wide supremacy territory, is about ${\rm Tr}(J_{22}J^{22})/2n$ away,
as measured by Kullback--Leibler. The power of the 5\% level
$Z_n^2=n(\hatt\gamma-\gamma_0)'(\hatt J^{22})^{-1}(\hatt\gamma-\gamma_0)$
test is about 13.3\% at the border for $q=2$ directions and
about 11.6\% at the border for $q=3$ directions, compared
to the previous 17.0\% power in the one-dimensional case. 
% where did I get my earlier figure 33.4\% from??? 
%	Power=Pr[chi^2_q(delta'(J^{22})^{-1}delta>upper 0.05 point]
% q=1:	Pr[chi^2_1(1) > 3.841] = 0.170.
% q=2:  Pr[chi^2_2(1) > 5.991] = 0.133. 
% Then 0.116, 0.105, 0.099, 0.094. q = 90 directions: 0.0587,
% very close to the narrow model ... 
So in a sense the tolerance region becomes more tightly concentrated
round the narrow model. But this tolerance ellipso\"\i d 
is the extremely cautious region where {\it all} estimands are 
better estimated using narrow methods. For a given estimand 
$\mu=\mu(\theta,\gamma)$, giving rise to a $q$-dimensional $b$ 
as in (3.2), there is a much wider area given by 
$(b'\delta)^2\le b'J^{22}b$
in which narrow estimation is more advantageous than wide estimation
(and only in the one-dimensional case does the estimand-dependent 
$b$ cancel out in this toleration criterion). 
This is the unbounded area between two hyperplanes lying 
tangentially to the cautious tolerance ellipso\"\i d. 
See K\aa resen (1992) for more details and for examples. 
We also mention that the discussion of 4D for the Akaike criterion
extends to give 
$$\eqalign{{\rm AIC}_{\rm wide}-{\rm AIC}_{\rm narr}
&\doteq n(\bar V_n-J_{21}J_{11}^{-1}\bar U_n)'J^{22}
	(\bar V_n-J_{21}J_{11}^{-1}\bar U_n)-2q \cr
&\arr_d\chi^2_q(\delta'(J^{22})^{-1}\delta)-2q. \cr}$$	
The Akaike method therefore chooses the narrow model with 
probability 0.843, 0.865, 0.888, 0.908 when $q=1,2,3,4$, 
if the narrow model is true. The probability of selecting
the narrow model at the border-line of the cautious tolerance region 
is correspondingly 0.653, 0.731, 0.788, 0.830. 
% General: AIC-difference goes to chi^2_q(delta'J^{22}delta)-2q.
% 	Narrow true, delta 0: 	Border-line, delta'J^{22}delta=1:
% 	Pr[chi^2_q(0)>2q]	Pr[chi^2_q(1)>2q]
% q=1   0.157			0.347
% q=2	0.135			0.269
% q=3 	0.112 			0.212
% q=4	0.092			0.170
% q=5	0.075			0.137
% q=6  	0.062			0.112
% q=90	0.05/10^6		0.09/10^6
%	--- It is shown in section 6(iii) that 
%	the Akaike method is inadmissible!  
%	Nils: satisfied? 
The probability that the Schwarz--Rissanen criterion mentioned in 4D
will choose the narrow model is about 
$\Pr\{\chi^2_q(\delta'(J^{22})^{-1}\delta\le q\log n\}$,
which goes (slowly) towards 1 for all local departures 
$\gamma_0+\delta/\sqrt{n}$. The trouble with these model choice 
criteria, in light of this discussion, is that they are 
too general and not dependent on the estimand, i.e.~the 
specific further use of the chosen model. 

Similarly there are generalisations of this section's results.
One can envisage useful generalisations of (5.4),
for example, where the final estimator gives weights to
the narrow model and to several wider alternative models,
with weights determined by the data, and possibly depending
on the estimand. 
Some results in these directions are in K\aa resen (1992). 
Statistics tradition does perhaps dictate this point of view,
with a classic null model and several possible departures from it,
but the problem can also be turned inside out, starting with
a wide a priori model for the data and then smoothing in several
directions downwards to narrower models of interest. 
The empirical Bayes ideas above should be of value if
these kind of questions are to be pursued. 
%	Inside and outside ... Diana Ross ... Two viewpoints ... 

\bigskip
{\bf 6. Grand comparison.}
Each estimator of $\mu_{\rm true}$ has a 
cousin that estimates $a$ in the $Z\sim \normal\{a,1\}$ situation,
and vice versa, by (5.8). Furthermore, 
the performance of one of them determines 
and is determined by the performance of the other one, by
the key correspondence (5.6)--(5.7). 
It is refreshing to judge a $\mu$-estimator 
by examining its $a$-estimator cousin. 
Here is a partial list of interesting estimators for $\mu_{\rm true}$,
following the various suggestions of Section 5, 
along with brief descriptions of their performance.  
%	--- See also Tables 1 and 2. Nils? --- 

(i) 
The narrow estimator $\hatt\mu_{\rm narr}$ 
has $c(z)\equiv0$ and $\hatt a(z)\equiv0$. This 
particular estimator of $a$ is fully confident that
$a$ is close to zero, and has risk $R_{\rm narr}(a)=a^2$.

(ii) 
The wide estimator $\hatt\mu_{\rm wide}$ 
on the other hand has $c(z)\equiv1$ and $\hatt a(z)=z$.
This conservative estimator has constant risk $R_{\rm wide}(a)=1$,
and is the unique admissible minimax estimator for $a$
when the parameter range is unrestricted. 
Note anew that the narrow is better than the wide when 
$|a|\le 1$. 

(iii) 
The if-else estimator (5.1), with $m^2$ instead of $1.645^2$
as cut-off point, 
has $c(z)=I\{|z|\ge m\}$, and corresponds to the $a$-estimator
$\hatt a(z)=zI\{|z|\ge m\}$. A determined mind finds 
$$\eqalign{
R_{\rm pre}(a)&=\int_{|z|\ge m}(z-a)^2\phi(z-a)\,\d z 
		+\int_{|z|\le m}(0-a)^2\phi(z-a)\,\d z \cr
	&=1+(a^2-1)\{\Phi(m+a)+\Phi(m-a)-1\}
		+(m+a)\phi(m+a)+(m-a)\phi(m-a). \cr}$$
The if-else with cut-off $m=1$,
seems overall to be preferable to both 
the one with $m=1.645$, corresponding to the 10\% level test,
and the one with $m=\sqrt{2}$, corresponding to the 
15.7\% level test that the Akaike criterion aims at, see 4D. 
The 10\% test is better in the vicinity of the narrow model,
for $|a|\le0.83$, but then becomes markedly worse than the former.
The pre-test estimators are not smooth enough to
be Bayes or extended Bayes, see (5.10). 
In particular such methods are not admissible,
i.e.~they can be improved upon uniformly in $a\,$! 
Note that $m=0$ and $m=\infty$ give back the wide and the narrow methods,
respectively. These extreme cases {\it are} admissible, however.

(iv) 
The linear combination estimator $\hatt\mu_{\rm lin}$ 
discussed in 5B has $c(z)=c$ and $\hatt a(z)=cz$. 
Its risk is $R_{\rm lin}(a)=c^2+(1-c)^2a^2$, 
which is unbounded when $|a|$ grows. 
These are proper Bayes solutions for $0\le c<1$ 
and admissible for $0\le c\le 1$.  

(v) 
The natural $\hatt\mu_{\rm eb}$ of (5.3) has $c(z)=z^2/(1+z^2)$,
and the correspondence to the empirical Bayes estimator
$\hatt a_{\rm eb}(z)=\{z^2/(1+z^2)\}z$ has been noted in 5F. 
One must compute 
$$R_{\rm eb}(a)=\E_a\Bigl({Z^2\over 1+Z^2}Z-a\Bigr)^2
	=\int\Bigl({z^3\over 1+z^2}-a\Bigr)^2\,\phi(z-a)\,\d z$$
%	=E\Bigl[{(a+X)^3\over 1+(a+X)^2}-a\Bigr]^2
by numerical integration. I can prove that $\hatt a_{\rm eb}$ is admissible.
This translates into an admissibility property for $\hatt\mu_{\rm eb}$.
I have also studied the similarly inspired 
$\hatt a(z)=\hatt q(z)z$, with $\hatt q(z)$ as in (5.11)
instead of $z^2/(z^2+1)$. These perform similarly.
The risk for $\hatt q(z)z$ starts at 0.37 for $a=0$, smaller
than 0.46 for $\hatt a_{\rm eb}(z)$, and stays better for $|a|\le 1.83$, 
after which $\hatt a_{\rm eb}(z)$ takes over. 
The maximum risk 1.476 for $\hatt q(z)z$
is higher than 1.252 for $\hatt a_{\rm eb}(z)$. 
The risk for $\hatt a_{\rm eb}(z)$ is less than
the crucial value 1 for $|a|\le1.40$. 
Overall one would argue that $\hatt a_{\rm eb}(z)$ 
is better than $\hatt q(z)z$;
see also the figure below. 

(vi) 
The restricted ML estimator (5.14) has risk function 
$$\eqalign{
R_{\rm res}(a)&=\Phi(m-a)+\Phi(m+a)-1-(m-a)\phi(m-a)-(m+a)\phi(m+a) \cr
     &\quad+(m-a)^2\{1-\Phi(m-a)\}+(m+a)^2\{1-\Phi(m+a)\}, \cr}$$
as some calculations show. 
This estimator is not smooth enough to be Bayes or extended
Bayes, and is like the if-else estimator not admissible. 
Its risk is satisfactory on $[-m,m]$ but ends up growing as 
$a^2$ outside it. 

\centerline{\includegraphics[scale=0.66]{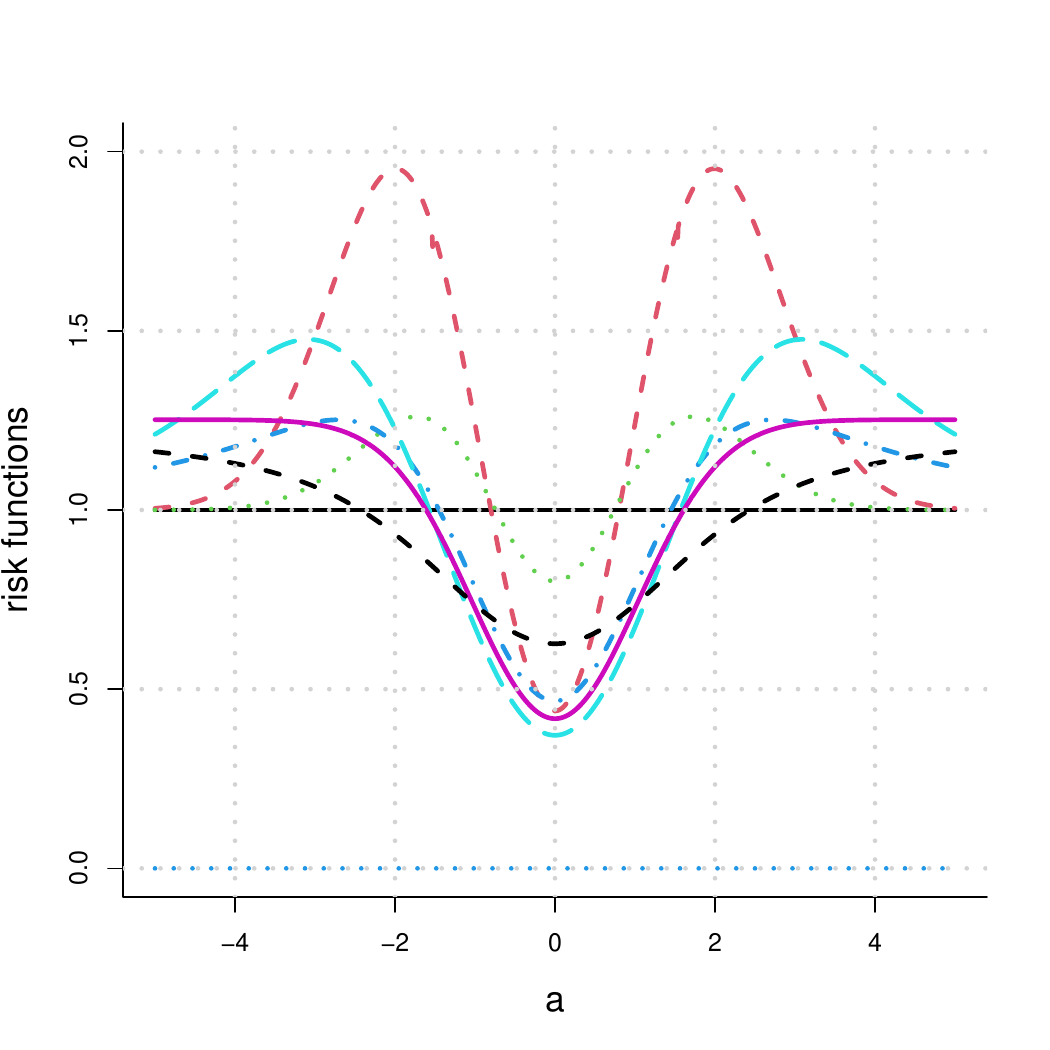}}

{{\medskip\narrower\sl\noindent\baselineskip11pt 
{\csc Figure.} 
Risk functions $R(a)$ are shown for seven procedures,
corresponding to seven choices of $c(Z_n)$ in (5.4),
as a function of $a=\delta/\kappa$, the normalised distance
from the narrow model. 
Risks for the wide and the narrow methods start at 1.00 and 0.00,
and are shown with dotted lines, 
as is the risk starting at 0.80 for the best pre-test method, 
with 1 rather than 1.645 in (5.1). 
The empirical Bayes methods
$\hatt a_{\rm eb}(Z)$ and $\hatt q(Z)Z$ start at 0.47 and 0.37.
Finally the Efron--Morris and arctan estimators,
both scaled to have the same maximum risk 1.252 
as has $\hatt a_{\rm eb}(Z)$, are those starting at 0.42 and 0.63. 
\medskip}}

(vii) 
The Efron--Morris estimator (5.15) has risk function 
$$R_{\rm em}(a)=1+m^2+(a^2-m^2-1)\{\Phi(m+a)+\Phi(m-a)-1\}
	-(m-a)\phi(m+a)-(m+a)\phi(m-a).$$
These increase with $|a|$, and rather rapidly,
from a small value at zero towards maximum risk $1+m^2$. 
The arctan-estimator (5.16) has also risk that increases 
in $|a|$ from $R_{\rm atan}(0)$ to $1+m^2$. It has higher risk than 
(5.15) has at $a=0$, but the risk climbs much more 
slowly towards $1+m^2$; see also the figure below. 
In particular an arctan-estimator can be better than
an Efron--Morris estimator on $[-5,5]$, say. 

% \smallskip
% \vskip8.5truecm
% \smallskip

% ! This is "plot7.com" of August 1991.
% ! It concerns different risk functions R(a) for a in my
% ! moderate miss paper, 
% ! for wide, narr, z^3/(1+z^2), q^(z)z, pretest(1), EM (.5017), atan (.5017).
% ! Could use \text='a'. 
% 
% device /portrait
% dimension /xlow=4.5 /ylow=17 /xhigh=16 /yhigh=25
% ! heading /text='Risk functions R(a) for seven estimators'
% data /file='today_data_17b.dat' 
% xaxis /low=0.0 /high=5.0 /interval=1.0 /tics=2 /variable=1 
%	/format=F /dec=0 /ypos=0.0
% ! unsuitable placing of: /text='a'; therefore dropped  
% yaxis /multi /low=0.0 /high=1.5 /interval=0.5 /tics=5 
%	/text='risk functions R(a)' /format=F /dec=1
% ! choose line=3 for wide, narr, pretest(1); line=1 for the four others.
% multi /variable=2 /line=3 
% ! multi /variable=3 /line=1 : modified version, column 9
% multi /variable=4 /line=1 
% multi /variable=5 /line=1 
% multi /variable=6 /line=3 
% multi /variable=7 /line=1 
% multi /variable=8 /line=1 
% multi /variable=9 /line=3 
% trunc /all=9.9999
% finish

% # This is com2 of gottseidank of May 1993, displaying risk functions
% # for my moderate mis-specificiation paper. 
% X <- matrix(scan("today_data_17b"),byrow=T,ncol=9)
% a <- X[,1]
% risks <- cbind(X[,2],X[,4],X[,5],X[,6],X[,7],X[,8],X[,9])
% # want columns 2 and 9 to be dotted and columns 4,5,6,7,8 to be lines! 
% # matplot(v,f,type="l") ... matlines?
% matplot(a,risks,type="l",lty=c(2,1,1,3,1,1,2),xlab="a=delta/kappa",ylab="risk % functions")
% # So narrow and wide are thinly dotted, pretest(1) is dotted,
% # and rest are smooth lines. 

(viii) 
The Bickel-inspired estimator (5.13) has acceptable
risk below 1 in an interval around 0, but the risk explodes when
$|a|$ grows. The same goes for the Bayes solution with a uniform
prior on $[-m,m]$. Its risk is 1 at 0 and at $m$ and below 1 in between,
but quickly explodes when $|a|$ grows outside the interval. 
Note that for $m$ large this solution becomes simply the wide solution. 

(ix) 
The `epsilon Bayes' methods described in (5.12) 
and the remarks following it have small risk for $|a|$ less than about 1,
but then become markedly worse than both 
$\hatt a_{\rm eb}(Z)$, $\hatt q(Z)Z$,
and pre-test estimators. The empirical epsilon Bayes method
with $\hatt\sigma^2=Z^2/\eps$ is not as good as the simple
specified one with $\sigma$ put equal to 3, for example. 
 
Let us compare 
$\hatt\mu_{\rm narr}$, 
$\hatt\mu_{\rm wide}$,
the if-else $\hatt\mu_{\rm pre}$ with $m=1$, 
and the mixture estimator $\hatt\mu_{\rm eb}$ of (5.3). 
The narrow estimator wins if $|a|\le 0.84$;
the mixture estimator wins when $|a|$ is between 0.84 and 1.45,
and finally the safest and wide estimator wins if $|a|$ exceeds 1.45.
While $\hatt\mu_{\rm narr}$ can misbehave significantly 
when $|\delta|\ge2.50\kappa$, say, 
$\hatt\mu_{\rm eb}$ always behaves wisely,
also in the $|\delta|>\kappa$ case,
and does not ever lose much to $\hatt\mu_{\rm wide}$.  
Its worst risk value is 1.252, at $|a|=2.70$, 
and when the narrow model is very wrong ($|\delta|$ is large)
$\hatt\mu_{\rm eb}$ becomes equivalent to $\hatt\mu_{\rm wide}$. 
In no region does $\hatt\mu_{\rm pre}$ win, but its risk function 
lies between the wide method's 1 and the mixture method's risk function, 
for $|a|>2.17$; see the figure. 

Based on these observations five of the more interesting 
estimators are singled out for display, 
in addition to the extreme basis choices `narrow' and `wide'. 
The five are the empirical Bayes versions 
$\hatt a_{\rm eb}(Z)$ and $\hatt q(Z)Z$; 
the pre-test strategy $\hatt a_{\rm pre}(Z)$ with 1 as cut-off, see 6(iii); 
the Efron--Morris (5.15) with $m=0.502$ chosen so as to 
get the same maximum risk 1.252 as $\hatt a_{\rm eb}(Z)$;
and the smoother arctan-estimator (5.16) with the same 
$m$ (and the same objective). 
The pre-test method with 1 as cut-off is about as good as 
these can be, but it is not as good as the others. 
It is included since versions of it are in frequent use. 
In this framework the Akaike method is one such. 

All in all the best choices appear to be the 
simple empirical Bayes, the Efron--Morris, and the arctan.
There are several other methods among those discussed 
that would make a good show on $[-5,5]$, say,
but with risks that explode for growing $|a|$. 
The $\hatt\mu_{\rm eb}$ of (5.3) in particular 
is a practical and satisfactory solution.
There is no artificial cut-off; its weight in favour of the wide
model is smoothly increasing from 0 to 1 with the test indicator $Z_n$; 
it behaves considerably better than the wide estimator 
in a reasonable neighbourhood of the narrow model; 
and its maximum risk is only $(1.119\,b\kappa)^2+\tau_0^2$,
compared to $(b\kappa)^2+\tau_0^2$ for the conservative wide method. 
The Efron--Morris and the arctan estimators have similar 
performances but require selection of a parameter, 
related to the trade-off between behaving well around zero and 
having a small maximum risk. 

The facts above are meant to summarise the main features 
of the various estimator performances, 
based on a thorough investigation and 
several days of conscientious staring at hundreds of risk functions. 
These were programmed via numerical integration when necessary. 
% Computer programs and risk tables for each of the estimator classes 
% discussed above are collected in Hjort (1991b), 
% which is available upon courteous request. 

To illustrate more concretely what these suggestions amount to, 
consider logistic regression as in Example F. 
If deviation from $\alpha+\beta x$ in direction of quadraticity
is suspected, use 
$$p^*(x)={1\over 1+Z_n^2}
	{\exp(\hatt\alpha_{\rm narr}+\hatt\beta_{\rm narr}x)
	\over 
	 1+\exp(\hatt\alpha_{\rm narr}+\hatt\beta_{\rm narr}x)}
 	+{Z_n^2\over 1+Z_n^2}
	{\exp(\hatt\alpha+\hatt\beta x+\hatt\gamma x^2)
	\over 
	 1+\exp(\hatt\alpha+\hatt\beta x+\hatt\gamma x^2)}, $$
where $Z_n=\sqrt{n}\hatt\gamma/\hatt\kappa$. 
Or replace the weights with $1-\hatt a(Z_n)/Z_n$ and $\hatt a(Z_n)/Z_n$,
with $\hatt a(Z_n)$ equal to the limited translation estimator (5.15)
or the arctan-estimator (5.16).

\bigskip
{\bf 7. Examples.}
We now provide answers to the questions asked in 
Examples A--G of the introduction!  

\subsection
{\csc Example A.} 
In the general two-parameter Weibull model,
parameterised as in (1.1), the score function becomes
$$\mtrix{{\gamma\over\theta}\{1-(\theta y)^\gamma\} \cr
	{1\over \gamma}\{1+\log(\theta y)^\gamma
		-(\theta y)^\gamma\log(\theta y)^\gamma\} \cr},$$
and clever computations involving the gamma function and its 
derivatives reveal the information matrix and its inverse to be 
$$J_{\rm gen}
	=\mtrix{\gamma^2/\theta^2 &(1-k)/\theta \cr
	          (1-k)/\theta     &c^2/\gamma^2 \cr},
	\quad
  J_{\rm gen}^{-1}={1\over \pi^2/6}
	\mtrix{c^2\theta^2/\gamma^2 &-(1-k)\theta \cr
		 -(1-k)\theta         &\gamma^2 \cr},$$
in which $k=0.577...$ is the Euler--Mascheroni constant and 
$c^2=\pi^2/6+(1-k)^2$. 
The null model corresponds to $\gamma=\gamma_0=1$. The $\kappa^2$ 
parameter is $6/\pi^2$, and we have reached the following conclusion:
For $|\gamma-1|\le\sqrt{6/\pi^2}/\sqrt{n}=0.779/\sqrt{n}$, 
estimation with $\mu(1/\bar Y,1)$ based on simple 
and narrow-minded exponentiality performs better than 
high-brow $\mu(\hatt\theta,\hatt\gamma)$;
and this is true regardless of the parameter $\mu$ to be estimated.  

In the language of 4F Weibull deviance from exponentiality 
has tolerance limit $d=J_{22}J^{22}=1+(1-k)^2/(\pi^2/6)=1.109$,
and $\rho^2$ of (4.3) is $(1-k)^2/\{(1-k)^2+\pi^2/6\}=0.098$.
It is instructive to compare these with corresponding values for
gamma distribution deviance from exponentiality. If
$f(y)=\{\theta^\gamma/\Gamma(\gamma)\}\,y^{\gamma-1}e^{-\theta y}$
is the gamma density, for which $\gamma_0=1$ gives back exponentiality,
then $\kappa^2=1/(\pi^2/6-1)$; estimation using $\mu(1/\bar Y,1)$
is more precise than $\mu(\hatt\theta,\hatt\gamma)$ provided
$|\gamma-1|\le1.245/\sqrt{n}$; $d$ is 2.551; and $\rho^2=6/\pi^2=0.608$.
This suggests that moderate gamma-ness is less critical than 
moderate Weibull-ness for standard methods based on exponentiality. 

\subsection
{\csc Example B.} 
The wide model has parameters $\xi$, $\sigma$, $m$. 
Let us reparameterise to $\gamma=1/m$, so that the density becomes
$$f(y,\xi,\sigma,\gamma)={c(\gamma)\over \sigma}
	\Bigl\{1+\gamma\Bigl({y-\xi\over \sigma}\Bigr)
			^2\Bigr\}^{-\{1/2+1/(2\gamma)\}}, \quad 
  c(\gamma)={\sqrt{\gamma}\over \sqrt{\pi}}
	{\Gamma(1/2+1/(2\gamma))\over \Gamma(1/(2\gamma))}.$$
Estimation of the model parameters must now be studied 
when $\gamma$ is small and nonnegative. 
This actually calls for special treatment since the null point
$\gamma=0$ is not an inner point, and $\hatt\gamma=0$,
or $\hatt m=\infty$, happens with positive probability.
Such a treatment is given in Hjort (1993),
and shows that if $\gamma\le0.686/\sqrt{n}$,
i.e.~if the degrees of freedom $m\ge1.458\sqrt{n}$, 
then $t$-ness doesn't matter, and {\it any} parameter 
$\mu=\mu(f)=\mu(\xi,\sigma,m)$ is better estimated in the ordinary,
simple, normality based way. A similar result is also 
proven there for regression models.

\subsection
{\csc Example C.} 
We generalise slightly and write 
the wide model as $Y_i\sim \normal\{\beta' x_i+\gamma c(x_i),\sigma^2\}$,
where $\beta$ and $x_i$ are $p$-dimensional vectors, and $c(x)$ is
some given scalar function. By computing log-derivatives and 
evaluating covariances one reaches 
$$J_{n,\rm wide}={1\over \sigma^2}
	\mtrix{2 &0 &0 \cr
		 0 &n^{-1}\sum_{i=1}^n x_ix_i' 
			&n^{-1}\sum_{i=1}^n x_ic(x_i) \cr
		 0 &n^{-1}\sum_{i=1}^n x_i'c(x_i) 
			&n^{-1}\sum_{i=1}^n c(x_i)^2 \cr},$$
from the definition in (3.5). It follows that 
$$\kappa^2=\sigma^2\times{\rm lower\ right\ element\ of\ }
  \mtrix{n^{-1}\sum_{i=1}^n x_ix_i' 
			&n^{-1}\sum_{i=1}^n x_ic(x_i) \cr
	   n^{-1}\sum_{i=1}^n x_i'c(x_i) 
			&n^{-1}\sum_{i=1}^n c(x_i)^2 \cr}^{-1}.$$
Assume, for a concrete example, that $x_i$ is one-dimensional 
and uniformly distributed over $[0,b]$, say $x_i=b{i\over n+1}$,
and that the wide model has $\alpha+\beta(x_i-\bar x)+\gamma(x_i-\bar x)^2$.
Then $\kappa\doteq \sqrt{80}\,\sigma/b^2$. Consequently,
dropping the quadratic term does not matter, and is actually
advantageous, for every estimator, 
provided $|\gamma|\le 8.94\,\sigma/b\sqrt{n}$. 
In many situations with moderate $n$ this will
indicate that it is best to keep the narrow model and 
avoid quadratic analysis. 

Similar analysis can be given for the case of a wide model
with an extra covariate, say $\normal\{\beta'x_i+\gamma z_i,\sigma^2\}$. 
The formulae above then hold with $z_i$ replacing $c(x_i)$. 
In the case of $z_i$'s distributed independently from the $x_i$'s
the narrow $x_i$ only model tolerates up to 
$|\gamma|\le(\sigma/\sigma_z)/\sqrt{n}$, where $\sigma_z^2$ is the 
variance of the $z_i$'s. 

\subsection 
{\csc Example D.} Again we are mildly general and 
write $Y_i\sim \normal\{\beta' x_i,\sigma^2(1+\gamma c(x_i))\}$
for the $p+2$ parameter variance heterogeneous model.  
It is not easy to put up simple expressions for 
the general information matrix, in the presence of $\gamma$,
but once more we are permitted to compute $J_{n,\rm wide}$ and
$J_{\rm wide}$ of (3.5) under the null model, that is,
when $\gamma=0$. Some calculations give 
$$J_{n,\rm wide}=\mtrix{
	2/\sigma^2 & 0 & \sigma^{-1}n^{-1}\sum_{i=1}^n c(x_i) \cr
	0 & \sigma^{-2}n^{-1}\sum_{i=1}^n x_ix_i' & 0 \cr
	\sigma^{-1}n^{-1}\sum_{i=1}^n c(x_i) & 0 
			& (2n)^{-1}\sum_{i=1}^n c(x_i)^2 \cr}.$$
Matters simplify and $1/\kappa^2$ is found to be 
$(2n)^{-1}\sum_{i=1}^n\{c(x_i)-\bar c\}^2$.
If once again $x_i$'s are distributed evenly on $[0,b]$,
and $c(x_i)=x_i$, 
then $\kappa\doteq \sqrt{24}/b$, and the criterion becomes
$|\gamma|\le 4.90/b\sqrt{n}$. In particular this shows that
the sophisticated variance heterogeneous approach, which 
uses the weighted least squares estimator
$$\hatt\beta_{\rm soph}=\sum_{i=1}^n{x_iy_i\over 1+\hatt\gamma c(x_i)}
	\Big/\sum_{i=1}^n{x_i^2\over 1+\hatt\gamma c(x_i)},$$
is inferior to the simpler solution, unless $|\gamma|$ is quite large.
It is of course the sampling variability present in the weights,
via the ML estimator $\hatt\gamma$, that makes $\hatt\beta_{\rm soph}$
inferior to ordinary $\hatt\beta_{\rm narr}$. 

\subsection
{\csc Example E.} 
Assume that a $h_\lambda$-transformation of 
$(Y_i-\beta' x_i)/\sigma$ is $\normal\{0,1\}$. 
When $h_\lambda(Z)$ is $\normal\{0,1\}$, then $Z$ has cumulative
$\Phi(z)^\lambda$ and density $\lambda\Phi(z)^{\lambda-1}\phi(z)$. 
Hence $Y_i$ has density
$$f(y_i,\sigma,\beta,\lambda\midd x_i)
	=\lambda\Phi\bigl((y_i-\beta' x_i)/\sigma\bigr)^{\lambda-1}\,
		\phi\bigl((y_i-\beta' x_i)/\sigma\bigr)/\sigma.$$
Is it now possible to evaluate partial derivatives w.r.t.~$\sigma$,
$\beta$, $\lambda$. Their null model versions, corresponding to
$\lambda=1$, become 
$(z_i^2-1)/\sigma$, $z_ix_i/\sigma$, $1+\log\Phi(z_i)$,
where $z_i=(y_i-\beta' x_i)/\sigma$. Formula (3.5) gives
the $(p+2)\times(p+2)$ matrix 
$$J_{n,\rm wide}=\mtrix{
	2/\sigma^2 &0 & b/\sigma \cr
	0 & \sigma^{-2}n^{-1}\sum_{i=1}^n x_ix_i' 
				&a\sigma^{-1}n^{-1}\sum_{i=1}^n x_i \cr
	b/\sigma & a\sigma^{-1}n^{-1}\sum_{i=1}^n x_i & 1 \cr},$$
in which $a=EN\log\Phi(N)=0.9032$ and $b=\E\{1+N^2\log\Phi(N)\}=-0.5956$
(computed by numerical integration). It follows that
$$1/\kappa^2=1-\half b^2
	-a^2\bar x'\Bigl({1\over n}\sum_{i=1}^n x_ix_i'\Bigr)^{-1}\bar x.$$
This $\kappa$ can be rather large, which in turn means that
standard regression copes well even if $\lambda$ 
differs quite a bit from 1. If only 
$|\lambda-1|\le \kappa/\sqrt{n}$, then standard regression 
methods work better than cumbersome ones employing 
a separate estimate for $\lambda$.

In the special case of a constant mean the tolerance 
limit against misspecification is very relaxed, 
with $\kappa=12.090$. In this case $V=1+\log\Phi(z)$ is
extremely well explained by $U=(z/\sigma,(z^2-1)/\sigma)$, 
with a maximal correlation of 0.993; see the discussion under 4F.
The classic $\normal\{\xi,\sigma^2\}$ can stand a good deal of 
misspecification w.r.t.~$\lambda$. --- In another  
special case, that of $\alpha+\beta(x_i-\bar x)+\sigma Z_i$,
$1/\kappa^2$ becomes $1-\half b^2$ and $\kappa$ becomes much 
smaller, namely 1.103. In the language of 4F 
the $n$ values of $V_i=1+\log\Phi(z_i)$ are now much less well
explained by the respective values of 
$U_i=\bigl(z_i,(x_i-\bar x)z_i,z_i^2-1\bigr)/\sigma$, 
and the standard regression model can only tolerate up to
$1.103/\sqrt{n}$ deviance from $\lambda=1$. 
%  	Grete, in Mathematica: a = 0.903197, b = -0.59564, 
%	kappa = 12.0898 in constant mean case;
%	kappa = 1.10256 in alpha + beat(x-xbar) case. Yes? 

\subsection
{\csc Example F.}
Write $p_i=p(x_i,\beta,\gamma)$, in which $\gamma=\gamma_0$ gives back
ordinary logistic regression, 
and write $p_i^0$ for $p(x_i,\beta_0,\gamma_0)$
at some target point $\beta_0$. 
It is not difficult to reach
$$J_{n,\rm wide}={1\over n}\sum_{i=1}^n
	{1\over p_i^0(1-p_i^0)}
	\mtrix{\dell p_i/\dell\beta \cr
		 \dell p_i/\dell\gamma \cr}
	\mtrix{\dell p_i/\dell\beta \cr
		 \dell p_i/\dell\gamma \cr}',$$
where the partial derivates are computed at the null point as usual.
Finding the tolerance limit $\kappa^2$ is achieved by 
computing this $(p+1)\times(p+1)$ matrix numerically, 
at the target point, which will typically be the estimated 
$\hatt\beta_{\rm narr}$ computed from ordinary analysis,
and then inverting it; $\kappa^2$ is found at the lower right corner.

In the two types of model departure discussed in Section 1, this
goes as follows. 
If the wide model says $\alpha+\beta(x_i-\bar x)+\gamma(x_i-\bar x)^2$,
then 
$$J_{n,\rm wide}={1\over n}\sum_{i=1}^n
	p_i^0(1-p_i^0)
	\mtrix{1     &t_i    &t_i^2 \cr
		 t_i   &t_i^2  &t_i^3 \cr
		 t_i^2 &t_i^3  &t_i^4 \cr},$$
where $t_i=x_i-\bar x$. In the case of (1.5), on the other hand,
involving a shape parameter $\eta$, 
$$J_{n,\rm wide}={1\over n}\sum_{i=1}^n
	{p_i^0\over 1-p_i^0}
	\mtrix{(1-p_i^0)^2 &(1-p_i^0)^2x_i &(1-p_i^0)\log p_i^0 \cr
		 (1-p_i^0)^2x_i &(1-p_i^0)^2x_i^2 
					&(1-p_i^0)\log p_i^0\,x_i \cr
		 (1-p_i^0)\log p_i^0 &(1-p_i^0)\log p_i^0\,x_i &
						       (\log p_i^0)^2 \cr}.$$
\subsection
{\csc Example G.}
Write $\sigma_1^2=\sigma^2$ and $\sigma_2^2=\sigma^2(1+\gamma)$.
Finding the $J_{\rm wide}$ matrix in the $(\xi_1,\xi_2,\sigma,\gamma)$ 
model is not difficult, and leads to 
$\kappa^2=2/\{r(1-r)\}$, where $r=m/(m+n)$. 
This means a tolerance level of $d=\kappa^2J_{22}=1/r=(m+n)/m$.
The simple equal variance model can tolerate 
$\gamma^2\le2(m+n)/mn$, which becomes $|\gamma|\le2/\sqrt{n}$ 
in the $m=n$ case. This is a fairly low tolerance limit,
and different variances qualifies as a dangerous departure 
from the narrow model.  

\subsection
{\csc Other examples.}
There is a large variety of other examples of 
common departures from standard models and that could be studied 
using our general methods and results. In each case one could 
compute the tolerance radius, one could speculate about 
robustness against the deviation in question in light of $d$ and 
$\rho$ of 4F, and one could implement the method of (5.3), for example. 
A partial list of such models and deviations is: 
(i) Typical i.i.d.~models against various forms of dependence.
(ii) Multinomial and log-linear models against higher order interactions. 
(iii) Analysis of variance models against interaction terms.
(iv) Analysis of variance models against different variances 
in different groups. 
(v) Regression models against presence of cross-terms. 
(vi) Time series models against higher order autoregression or 
moving average.
(vii) Typical i.i.d.~models for discrete variables against Markov
dependence.
(viii) Markov chain models against second order Markovness. 
Some results and examples for this and the previous situation
are in Fenstad (1992). 
(ix) Models with normal errors against contamination of gross
errors.  
(x) Traditional homogeneous models in survival analysis against 
heterogeneous frailness of individuals. 
(xi) Normal class densities with common covariance matrix 
against moderately different covariance matrices in discriminant analysis. 
% 	Anne-Marie; Nils; Kjetil; Cristina; ... 
%
%\bigskip
%{\bf Acknowledgement.} 
%This work was ignited during a talk on simulation experiments 
%on misspecified logistic regression models, held by Petter Laake 
%last Thursday. 

\bigskip
{\bf Acknowledgements.} 
I have had fruitful discussions on aspects of this article with
my graduate students Anne Marie Fenstad and Kjetil K\aa resen. 
% for whom? 
Comments from Peter Bickel, Kjell Doksum and David Pollard 
have been helpful. A part of this work was completed while 
visiting University of Oxford with partial support 
from the Royal Norwegian Research Council. 

\bigskip
\parindent0pt
\parskip3pt
\baselineskip11pt

\centerline{\bf References}

\medskip 
\ref{%
Berger, J.O. (1982).
Estimation in continuous exponential families:
Bayesian estimation subject to risk restrictions 
and inadmissibility results.
In {\sl Statistical Decision Theory and Related Topics III},
editors Berger and Gupta, 109--141. 
Academic Press, New York.} 

\ref{%
Bickel, P.J. (1981). 
Minimax estimation of the mean of a normal
distribution when the parameter space is restricted.
{\sl Annals of Statistics}~{\bf 9}, 1301--1309.}

\ref{%
Bickel, P.J. (1983). 
Minimax estimation of the mean of a normal distribution 
subject to doing well at a point.
{\sl Recent Advances in Statistics,
Festschrift for Herman Chernoff},
editors Rizvi and Siegmund, 511--528.
Academic Press, New York.}
 
\ref{%
Bickel, P.J. (1984). 
Parametric robustness: small biases can be worthwhile.
{\sl Annals of Statistics}~{\bf 12}, 864--879.}

\ref{%
Efron, B.~and Morris, C. (1971).
Limiting the risk of Bayes and empirical Bayes estimators.
{\sl Journal of the American Statistical Association}~{\bf 66}, 807--815.}

\ref{%
Fenstad, A.M. (1992).
How much dependence can the independence assumption tolerate?
Graduate thesis (in Norwegian), 
Department of Mathematics and Statistics, University of Oslo.}

\ref{%
Hjort, N.L. (1986a). 
{\sl Theory of Statistical Symbol Recognition.}
Research monograph, Norwegian Computing Centre, Oslo.}
 
\ref{%
Hjort, N.L. (1986b). 
Bayes estimators and asymptotic efficiency in parametric
counting process models.
{\sl Scandinavian Journal of Statistics}~{\bf 13}, 63--85.}

\ref{%
Hjort, N.L. (1993).
The exact amount of t-ness that the normal model can tolerate.
% Technical report, University of Oslo;
Submitted for publication.}

% \ref{%
% Hjort, N.L. (1991b).
% Computer programs and risk functions for estimators for a normal mean.
% Technical report, University of Oslo;
% available upon courteous request.}

\ref{%
K\aa resen, K. (1992).
Parametric estimation: Choosing between narrow and wide models.
Graduate thesis, Department of Mathematics and Statistics,
University of Oslo.}
% Also issued as Statistical Research Report No.~4/92.

\ref{%
Lehmann, E.L. (1983). 
{\sl Theory of Point Estimation.} Wiley, Singapore.} 

% Efron--Morris? 
% Bickel and Doksum, JASA, on Box--Cox; Replies \& Rebuttals (?).
% Brown, L.D. (1971).
% Admissible estimators, recurrent diffusions, and insoluble
% boundary value problems.
% {\sl Ann.~Math.~Statist.}~{\bf 42}, 855--903.
%
% Casella, G.~and Strawderman, W. (1981).
% Estimating a bounded normal mean.
% {\sl Ann.~Statist.}~{\bf 9}, 868--876. 

\bye